\documentclass[11pt]{article}
\usepackage[utf8]{inputenc}
\usepackage[T1]{fontenc}
\pdfoutput = 1

\usepackage{amsmath}
\usepackage{amssymb}
\usepackage{bbm}
\usepackage{bbold}
\usepackage{braket}
\usepackage{caption}
\usepackage{color}
    \definecolor{darkgreen}{rgb}{0,0.5,0}
    \definecolor{darkred}{rgb}{0.5,0,0}
    \definecolor{darkblue}{rgb}{0,0,0.6}
    \definecolor{purple}{rgb}{0.4,.2,0.7}
\usepackage[mathcal]{eucal}
\usepackage{float}
\usepackage{graphicx}
\usepackage[hyperfootnotes = true, colorlinks = true, linkcolor = darkblue, citecolor = purple]{hyperref}
\usepackage{mathabx}
\usepackage{mathtools}
\usepackage{pdfsync}
\usepackage{slashed}
\usepackage[normalem]{ulem}
\usepackage{upgreek}
\usepackage{url}

\allowdisplaybreaks

\usepackage[margin = 2.2cm]{geometry}
\usepackage[ragged]{footmisc}
\renewcommand{\d}{\mathrm{d}}
\renewcommand{\i}{\mathrm{i}}

\DeclareMathOperator{\Tr}{Tr}

\DeclareMathOperator{\vol}{vol}
\DeclareMathOperator{\arccosh}{arccosh}
\newcommand{\veff}{V_\text{eff}}

\newcommand\stac[2]{\genfrac{}{}{0pt}{}{#1}{#2}}
\newcommand{\ev}[1]{\left\langle #1 \right\rangle}
\newcommand{\cC}{\mathcal{C}}

\numberwithin{equation}{section}

\begin{document}

\thispagestyle{empty}
\begin{center}
    ~\vspace{5mm}
    
    {\Large \bf 

        Complex eigenvalue instantons and the Fredholm determinant expansion in the Gross-Witten-Wadia model
    
    }
    
    \vspace{0.4in}
    
    {\bf Dan Stefan Eniceicu,$^1$ Raghu Mahajan,$^1$ and Chitraang Murdia.$^{2,3}$}

    \vspace{0.4in}

    $^1$ Department of Physics, Stanford University, Stanford, CA 94305-4060, USA \vskip1ex
    $^2$ Berkeley Center for Theoretical Physics, Department of Physics, University of California, Berkeley, CA 94720, USA \vskip1ex
    $^3$ Theoretical Physics Group, Lawrence Berkeley National Laboratory, Berkeley, CA 94720, USA
    \vspace{0.1in}
    
    {\tt eniceicu@stanford.edu, raghumahajan@stanford.edu, murdia@berkeley.edu}
\end{center}

\vspace{0.4in}

\begin{abstract}

We study the leading nonperturbative corrections to the strong-coupling (ungapped) phase of the Gross-Witten-Wadia (GWW) integral over unitary matrices, to one-loop order.
We compute these corrections directly in terms of eigenvalue tunneling in a holomorphic presentation of the integral over eigenvalues.
The leading nonperturbative contribution to the partition function comes from a pair of complex eigenvalue instantons.
We show that these are in fact ``ghost instantons'', which are extrema of the one-eigenvalue effective potential on the ``unphysical sheet'' of the spectral curve and have been discussed in detail recently by Mari\~no, Schiappa, and Schwick.
Further, we discuss the relationship of these instantons to the Fredholm determinant expansion of the unitary matrix integral, which has recently become an object of interest in the computations of BPS indices of supersymmetric gauge theories and black holes.
We find that, after taking the 't Hooft limit, the first correction given by the Fredholm determinant expansion of the GWW integral agrees precisely with the leading nonperturbative correction, to one-loop order.

\end{abstract}

\pagebreak

\tableofcontents

\section{Introduction and setup}
In this short note, we study the Gross-Witten-Wadia (GWW) integral \cite{Gross:1980he, Wadia:1980cp, Wadia:2012fr}
\begin{align}
    Z(N,t) := \int \frac{\d U}{\vol U(N)} \, \exp\left(
    \frac{N}{2t} \left(\Tr U + \Tr U^{-1} \right)
    \right)\, .
    \label{defznt}
\end{align}
The integral is over $N \times N$ unitary matrices $U$.
The original papers \cite{Gross:1980he,Wadia:1980cp,Wadia:2012fr} arrived at this integral via studies of 2D lattice gauge theory with gauge group $U(N)$.
We are interested in the 't Hooft large-$N$ limit, where the 't Hooft coupling $t$ is held fixed as $N \to \infty$.
For later use, we will also define
\begin{align}
    \tau := \frac{N}{2t}\, , \quad Z_N(\tau) := Z(N,t)\, .
    \label{auxdef}
\end{align}

With 't Hooft scaling, this integral exhibits a third-order phase transition at $t=1$ \cite{Gross:1980he,Wadia:1980cp,Wadia:2012fr}.
Performing the change of variables from the matrix $U$ to its eigenvalues \cite{Eynard:2015aea}, we can rewrite $Z(N,t)$ as 
\begin{align}
    Z(N,t) = \frac{1}{N!} \int \prod_{i=1}^N \frac{\d \theta_i}{2\pi} \,
    \prod_{j<k} \vert e^{\i \theta_j} - e^{\i \theta_k} \vert^2  \,
    \exp \left( \frac{N}{t} \sum_{i=1}^N \cos \theta_i \right)\, .
    \label{znttheta}
\end{align}
The basic observation is that when $t$ is small, the potential $-\frac{1}{t} \cos\theta$ is steep and confines the eigenvalues to lie in a small interval, symmetric around $\theta = 0$.
As $t$ increases, the potential becomes weaker and the width of the eigenvalue distribution increases because of the Vandermonde repulsion.
At $t=1$, the width of the eigenvalue distribution becomes $2\pi$ and the eigenvalues cover the whole circle.
This signals a phase transition to the ungapped phase for $t>1$.
As $t\to \infty$, the potential vanishes; this reduces the problem to the pure Haar measure over the unitary group, and the eigenvalues are uniformly distributed over $\theta$.
In the strong-coupling phase ($t>1$), the perturbative approximation to various quantities can be obtained by expanding the exponential in the integrand $\exp (\tau (\Tr U + \Tr U^{-1}))$ in powers of $\tau$, and using the Haar integrals computed in \cite{diaconis1994eigenvalues}, without worrying about finite-$N$ trace relations.

Let us say a few more words about this phase transition.
Large-$N$ matrix integrals are famously known to possess nonperturbative effects due to one-eigenvalue instantons \cite{David:1990sk, Shenker:1990uf, Ginsparg:1991ws}. These are saddles of the matrix integral in the large-$N$ limit where one eigenvalue is removed from the cut (i.e. the original support of the eigenvalue density) and moved to an extremum of the \emph{effective} potential (which is a combination of the bare potential and the Vandermonde repulsion \cite{Eynard:2015aea, DiFrancesco:1993cyw}).
In the gapped, weak-coupling phase ($t<1$), there is an eigenvalue instanton at $\theta = \pi$. 
It is a very interesting fact that the action of this instanton goes to zero continuously at the phase transition point, thus becoming unsuppressed \cite{Marino:2008ya}.
In the ungapped, strong-coupling phase ($t>1$), the eigenvalue instantons are qualitatively different and they are located at complex values of $\theta$ \cite{Buividovich:2015oju}. 

The GWW integral exhibits a double-scaling limit, in which one zooms into the $t\approx 1$ region.
In this double-scaling limit, the solution to the GWW integral is described by the Hastings-McLeod solution \cite{hastings1980boundary} to the Painlev\'e-II equation \cite{Marino:2008ya}.
In this limit, it is also equal to the partition function of a particular minimal superstring model \cite{Klebanov:2003wg}.

Our specific interest is in the instanton contributions to $Z(N,t)$, in the ungapped, or the strongly-coupled phase $t>1$, outside the double-scaling limit (but still staying within the 't Hooft limit).
The genus expansion in this phase only has a planar contribution; all the $1/N^2$ corrections to the free energy vanish.
However, instanton corrections still exist.
This fact was appreciated long ago \cite{Goldschmidt:1979hq} and was studied in detail by Mari\~no \cite{Marino:2008ya} using transseries solutions to the ``string equation'' both in and away from the double-scaling limit.
The explicit form for the leading instanton correction in the $t>1$ phase was written down in \cite{Ahmed:2017lhl} using a relationship between the partition function and $\ev{\det U}$.
It reads
\begin{align}
    \frac{Z(N,t)}{\exp(N^2/4t^2)} &= 1 -
    e^{-2 N S_\text{strong}(t)} \,
    \frac{1}{N} \frac{t}{8\pi (t^2 - 1)^{3/2}} \left( 1 + O(N^{-1}) \right)
    + O(e^{-4 N S_\text{strong}(t)}) \, ,
    \label{main-target} \\
    S_\text{strong}(t) &=  \arccosh(t) - \sqrt{1-t^{-2}}\, , \quad \quad (t>1)\, .\label{sstrong}
\end{align}
The denominator on the left side is the perturbative contribution to $Z(N,t)$, as reviewed below.

One of the goals of the present work is to give a derivation of the result \eqref{main-target} via a direct analysis of the saddle points of the eigenvalue integral (\ref{znttheta}). The lack of such a derivation until now constitutes a fundamental gap in the literature given that the original one-eigenvalue instanton computations in the early '90s were done using this approach \cite{David:1990sk} which has since been adopted as the standard technique for calculating nonperturbative effects in matrix integrals. The hindrance people faced in adapting this computation to the GWW integral is that the instanton action predicted by standard eigenvalue tunneling has the wrong sign and would therefore incorrectly suggest that the nonperturbative corrections should be exponentially enhanced, rather than suppressed in the GWW integral \cite{Buividovich:2015oju}. We overcome this hindrance by showing that the exponentially-suppressed contribution in \eqref{main-target} is, in fact, attributed to the recently discovered ``ghost instantons''.\footnote{
Ghost instantons are nonperturbative configurations with opposite actions compared to those of regular instantons. 
The existence of these nonperturbative sectors was first predicted by studying the resurgent properties of the differential equation governing the specific heat of the $(2,3)$ minimal string \cite{Garoufalidis:2010ya} (see also \cite{Klemm:2010tm}). 
They were subsequently studied in \cite{Aniceto:2011nu,Schiappa:2013opa,Gregori:2021tvs,Baldino:2022aqm}, and were successfully explained in the context of Hermitian matrix integrals in terms of anti-eigenvalue tunneling in \cite{Marino:2022rpz}. 
Finally, they were given a physical interpretation in minimal string theory as ghost D-branes in \cite{Schiappa:2023ned}. 
See \cite{Gu:2022sqc,Gu:2022fss,Gu:2023mgf,Eynard:2023qdr} for subsequent applications.}
As explained in \cite{Marino:2022rpz}, in large-$N$ matrix integrals with a two-sheeted spectral curve, each eigenvalue instanton with action $S$ should have a partner ghost instanton, whose action is $-S$ and represents eigenvalue tunneling to the unphysical sheet. 
It is very interesting that the simple and heavily studied GWW unitary matrix integral contains a directly measurable ghost instanton effect. 
We also convincingly show that the leading nonperturbative contribution is a two-instanton effect.
Furthermore, the present direct eigenvalue/ghost instanton approach to deriving nonperturbative corrections to the asymptotics of matrix integrals has the advantage of being easily generalized to unitary matrix integrals with more complicated potentials, whereas the differential equation approach of \cite{Ahmed:2017lhl} is specific to the GWW potential.

The second goal of this work is to relate eigenvalue instantons to the Fredholm determinant expansion of $Z(N,t)$ \cite{geronimo1979scattering, borodin2000fredholm}.
There is an exact formula for $Z(N,t)$ in terms of a Toeplitz determinant \cite{Goldschmidt:1979hq}
\begin{align}
    Z(N,t) = \det \bigg[ 
    I_{k-l}(N/t) \bigg]_{k,l = 1,\ldots, N} \, .
    \label{toeplitz}
\end{align}
Here $I_n(x)$ denotes the $n$th Bessel-I function.
The theorem of \cite{geronimo1979scattering, borodin2000fredholm} relates this Toeplitz determinant to a Fredholm determinant.
Recently, this Fredholm determinant expansion has attracted interest because of its role in supplying finite-$N$ corrections to the index of supersymmetric black holes in AdS, and of BPS states in supersymmetric gauge theories \cite{Murthy:2022ien}.
The index is computed by a unitary integral with a double-trace potential, which, after a Hubbard-Stratonovich transformation, is related to a GWW-type integral with an infinite number of single-trace terms in the potential.\footnote{These couplings are controlled by a few chemical potentials that enter in the definition of the symmetry-resolved index.}
The $m$th term in this expansion of the index is of order $e^{- c_1 m N}$, and is related to contributions to the index coming from $m$ giant gravitons or wrapped D-branes \cite{Gaiotto:2021xce, Lee:2022vig, Arai:2019aou,Arai:2019wgv,Arai:2019xmp,Arai:2020qaj,Arai:2020uwd,Fujiwara:2021xgu,Imamura:2021ytr, Kimura:2021lrc, Fujiwara:2023bdc}.\footnote{The Fredholm determinant expansion is related to, but not exactly the same as the giant-graviton expansion \cite{Liu:2022olj}. However, it was shown that it is possible to extract the terms of the giant-graviton expansion from those of the Fredholm determinant expansion at least in the single-fugacity case \cite{Eniceicu:2023uvd}. For important clarifications regarding the subtleties surrounding wall-crossing in the multiple-fugacity case see \cite{Beccaria:2023zjw}.}
This observation provides the underlying intuition for why the Fredholm determinant expansion of the GWW integral itself should be related to the eigenvalue instanton expansion, which also supplies corrections of order $e^{-c_2 N}$.
We will show that this intuition is precise. 
The Fredholm determinant expansion is usually written in terms of the quantity $\tau$ (which is why we introduced the auxiliary definitions (\ref{auxdef})), and we will show that by taking the 't Hooft limit, the first nontrivial term in the Fredholm determinant series agrees with \eqref{main-target}.

The structure of this paper is as follows.
In section \ref{sec:eiginst}, we demonstrate that a direct analysis of eigenvalue instantons in the strongly-coupled phase of the GWW integral reproduces the result \eqref{main-target}.
We demonstrate how eigenvalue instantons on the ``unphysical sheet'' arise in this context and are the ones that actually contribute.
In section \ref{sec:fredholm}, we demonstrate that the first nontrivial term in the Fredholm determinant expansion of $Z(N,t)$ also reproduces \eqref{main-target} in the 't Hooft limit.
In section \ref{sec:summary}, we summarize the results and discuss some open problems.
Appendix \ref{apperturbative} contains some results for the perturbative correlators that we need.
In appendix \ref{appdet}, we analyze the quantity $\ev{\det U}$ and directly compute the first nonperturbative contribution to it using the integral over eigenvalues. 
The contribution comes from a single ghost instanton.
In appendix \ref{oneeigenvalueinstanton}, we analyze the putative one-instanton contribution to the partition function and show how it eventually becomes a two-instanton contribution, giving another derivation of (\ref{main-target}).
In appendix \ref{supermatrixapp}, we analyze the problem from the perspective of tunneling anti-eigenvalues of an associated supermatrix integral. 
Finally, in appendix \ref{appC}, we extend the analysis of the leading nonperturbative effects due to regular eigenvalue instantons as well as ghost instantons to the case of the strong-coupling phase of matrix integrals with general (higher-order) single-trace potentials.
For this more general class of potentials, the differential equation technique of \cite{Ahmed:2017lhl} is not available for the precise computation of instanton corrections, and thus our techniques provide new results in these cases.

\section{Complex eigenvalue instantons in the strong-coupling phase}
\label{sec:eiginst}

In this section, we will reproduce (\ref{main-target}) by a direct analysis of the eigenvalue instantons in the integral (\ref{defznt}), or equivalently, in (\ref{znttheta}).
Previously, this result was obtained in \cite{Ahmed:2017lhl} by an analysis involving the expectation value of $\det U$ and a certain nonlinear differential equation that it satisfies.

It will help us to cast (\ref{znttheta}) into a holomorphic form; see, for example, Section 1.2.3 of \cite{Eynard:2015aea}.
Let $z_j := e^{\i \theta_j}$ and note that
\begin{align}
    \vert e^{\i \theta_j} - e^{\i \theta_k} \vert^2 &= - \frac{1}{z_j z_k} (z_j - z_k)^2 \, , \quad 
    \frac{\d\theta_i}{2\pi} = \frac{\d z_i}{2\pi \i z_i}\, .
\end{align}
Introducing the potential
\begin{align}
    V(z) = - \frac{1}{2}(z + z^{-1})\, ,
\end{align}
we can rewrite (\ref{znttheta}) as
\begin{align}
    Z(N,t) &= (-1)^{\frac{1}{2}N(N-1)}\frac{1}{N!} \int 
    \prod_{i=1}^N \frac{\d z_i}{2\pi \i z_i^N} \prod_{j<k} (z_j - z_k)^2
    \exp \left( - \frac{N}{t} \sum_{i=1}^N V(z_i) \right) \label{zntholomorphic1} \\
    &=(-1)^{\frac{1}{2}N(N-1)} \, \i^{-N}\frac{1}{N!} \int 
    \prod_{i=1}^N \frac{\d z_i}{2\pi} \prod_{j<k} (z_j - z_k)^2
    \exp \left( \frac{N}{2t} \sum_{i=1}^N (z_i + z_i^{-1} - 2t \log z_i) \right) \, .
    \label{zntholomorphic}
\end{align}
The advantage of this last form is that the integral over eigenvalues looks locally like one that would be obtained from a matrix integral over Hermitian matrices.
The cost we pay is a logarithmic term in the potential.
So, in principle, we can use the results of \cite{Marino:2008vx, Schiappa:2013opa, Eniceicu:2022dru} expressing the one-loop instanton contribution in terms of the perturbative data around the leading one-cut saddle point.
In fact, as mentioned in the introduction, we will need the corresponding formulas for the ghost instantons developed in \cite{Marino:2022rpz}.
However, the presence of a logarithmic term in the potential can cause ambiguities because of the multi-valuedness of the logarithm in the complex plane.
Furthermore, since the eigenvalue distribution in the strong-coupling phase covers the entire unit circle, the spectral curve consists of two disconnected sheets. 
Because of these differences, we will derive all the needed formulas from scratch, and for this we will find it useful to use the form of the integral in (\ref{zntholomorphic1}) in which the integrand does not have any branch cuts.

\subsection{The perturbative expansion in the strong-coupling phase}
In this subsection, as a warm-up and review, we compute the perturbative approximation to various important quantities. 
These ingredients will be used in the next subsection to obtain the instanton correction to one-loop order.

In the strong-coupling phase, the perturbative expansions for various quantities can be obtained by directly expanding the exponential in the defining integral (\ref{defznt}) and using the following result of Diaconis and Shahshahani (Theorem 2 of \cite{diaconis1994eigenvalues}):
\begin{align}
    \int \frac{\d U}{\vol U(N)} \, \prod_{j=1}^k 
    (\Tr U^j)^{a_j} (\Tr U^{-j})^{b_j} &= \prod_{j=1}^k j^{a_j} (a_j)!\, \delta_{a_j, b_j} 
    \quad\quad \text{if } N \geq \sum_{j=1}^k j \, a_j\, .
    \label{diacshah}
\end{align}
The point is that at large $N$ and in the strong-coupling phase, the perturbative quantities can be computed by ignoring the constraint $N \geq \sum_{j=1}^k j \, a_j$ in this result.

Let us see an example of this result in action and compute the perturbative approximation to $Z(N,t)$ itself, by directly expanding the exponential in (\ref{defznt}):
\begin{align}
    Z(N,t) = \int \frac{\d U}{\vol U(N)} \sum_{j=0}^\infty 
    \frac{1}{j!} \left( \frac{N}{2t} \right)^j (\Tr U + \Tr U^{-1})^j \, .
\end{align}
Because of the Kronecker deltas in (\ref{diacshah}), we will only get a nonzero contribution if $j$ is even and, furthermore, when expanding $(\Tr U + \Tr U^{-1})^j$ via the binomial theorem, we only keep the middle term. 
Thus, the perturbative large-$N$ partition function in the strong-coupling phase is
\begin{align}
    Z^{(0)}(N,t) &= \int \frac{\d U}{\vol U(N)} \sum_{k=0}^\infty 
    \frac{1}{(2k)!} \left( \frac{N}{2t} \right)^{2k} 
    \frac{(2k)!}{(k!)^2} (\Tr U)^k (\Tr U^{-1})^k \\
    &= \sum_{k=0}^\infty \frac{1}{k!^2} \left( \frac{N}{2t} \right)^{2k} \times k!
    = \exp \left( \frac{N^2}{4t^2}\, \right) .
    \label{zntperturbative}
\end{align}
We used the integral (\ref{diacshah}) in the second line.
This result agrees with the known planar free energy of the GWW model \cite{Gross:1980he, Wadia:1980cp, Wadia:2012fr, Marino:2008ya}.

Next, we consider the computation of the resolvent. 
We would like to define the resolvent as $R(z) := \frac{1}{N} \left\langle \Tr \frac{1}{z-U} \right\rangle$.
However, it is well-known that the eigenvalue distribution in the strong-coupling phase of the GWW integral covers the whole circle and so, in fact, we will have two resolvents:
\begin{align}
    R^+(z) &:= 
    \frac{1}{N} \left\langle \Tr \frac{1}{z-U} \right\rangle \, ,
    \quad \quad\text{for } \vert z \vert > 1 \label{rplusdef} \, , \quad \text{and} \\
    R^-(z) &:= \frac{1}{N} \left\langle \Tr \frac{1}{z-U} \right\rangle \, ,
    \quad \quad \text{for } \vert z \vert < 1 \, . \label{rminusdef}
\end{align}
These are the defining equations, but $R^+(z)$ can be analytically continued from the outside to the inside of the unit disk, and $R^-(z)$ can be analytically continued from the inside to the outside of the unit disk.

For $R^+(z)$, by expanding $\frac{1}{z-U} = \sum_{k=1}^\infty z^{-k} U^{k-1}$ as appropriate for $\vert z \vert > 1$, expanding the exponential in the GWW measure (\ref{defznt}) as in the computation of $Z(N,t)$, and using the result (\ref{diacshah}) we obtain
\begin{align}
    R^+(z) = \frac{1}{z} + \frac{1}{z^2} \frac{1}{2t} \, .
\end{align}
The asymptotic behavior $R^+(z) \to \frac{1}{z}$ as $z\to \infty$ is needed for the correct normalization of the density of states.
The computation of $R^-(z)$ is similar, except we need to expand $\frac{1}{z-U} = - \sum_{k=1}^\infty U^{-k} z^{k-1} $ as appropriate for $\vert z \vert < 1$.
The result is
\begin{align}
    R^-(z) = - \frac{1}{2t}\, . 
\end{align}
We note that as $t \to \infty$, these results reduce to the known resolvents for the pure Haar ensemble.\footnote{See, for example, the discussion around Eq.~(3.2.52) of \cite{Eynard:2015aea}.}
Furthermore, these results allow us to extract the density of states via
\begin{align}
    \rho(\theta) = \frac{R^+(z) - R^-(z)}{2\pi \i} \frac{\d z}{\d \theta} 
    = \frac{1}{2\pi}\left(1 + \frac{1}{t} \cos\theta \right)\, ,
    \quad \theta \in [-\pi,\pi)\, .
\end{align}
Again, this result for the density of states is well-known \cite{Gross:1980he, Wadia:1980cp, Wadia:2012fr, Marino:2008ya}.
Note that this density does not make sense in the weak-coupling phase $t < 1$ since the expression would not be positive for all $\theta$.

\subsection{Instanton contributions in the strong-coupling phase}
Now that we know the resolvents and the density of states, we can compute the derivative of the one-eigenvalue effective potential. 
The one-eigenvalue effective potential is the sum of the bare external potential and a logarithmic two-body repulsion term caused by the Vandermonde determinant, $\veff(z) = V(z) + t \log z - \frac{2 t}{N}\left\langle \Tr \log (z - U)\right\rangle$.
We are following the conventions of \cite{Marino:2008ya, Eniceicu:2022dru}.
The second term $t/z$ is due to the logarithmic term in the bare potential appearing in (\ref{zntholomorphic}).
Because of the same reasons as explained in the computation of the resolvent, in fact we need two analytic functions $\veff^+(z)$ and $\veff^-(z)$ defined outside and inside the unit circle, respectively.
We  have
\begin{align}
    \frac{\d}{\d z}\veff^+(z) &= -\frac{1}{2} \left(1 - z^{-2} \right) 
    + \frac{t}{z} - 2t \, R^+(z)
    = - \frac{1}{2} - \frac{1}{2z^2} - \frac{t}{z}\, ,\\
    \frac{\d}{\d z}\veff^-(z) &= -\frac{1}{2} \left(1 - z^{-2} \right) 
    + \frac{t}{z} - 2t \, R^-(z)
    = \frac{1}{2} + \frac{1}{2z^2} + \frac{t}{z}\, .
\end{align}
We note that the expression for $\frac{\d}{\d z}\veff^-(z)$ is simply the negative of the expression for $\frac{\d}{\d z}\veff^+(z)$.
The two common zeroes of these functions are the locations of the eigenvalue instantons:
\begin{align}
    z_1^\star = -t + \sqrt{t^2 - 1} \, , \quad 
    z_2^\star = -t - \sqrt{t^2 - 1} \, .
    \label{zstars}
\end{align}
We note that both $z_1^\star$ and $z_2^\star$ are real and negative, with $z_1^\star$ being inside the unit circle and $z_2^\star$ being outside the unit circle.
These locations were also found numerically in \cite{Buividovich:2015oju}.

\begin{figure}
    \centering
    \includegraphics[width = 0.65\textwidth]{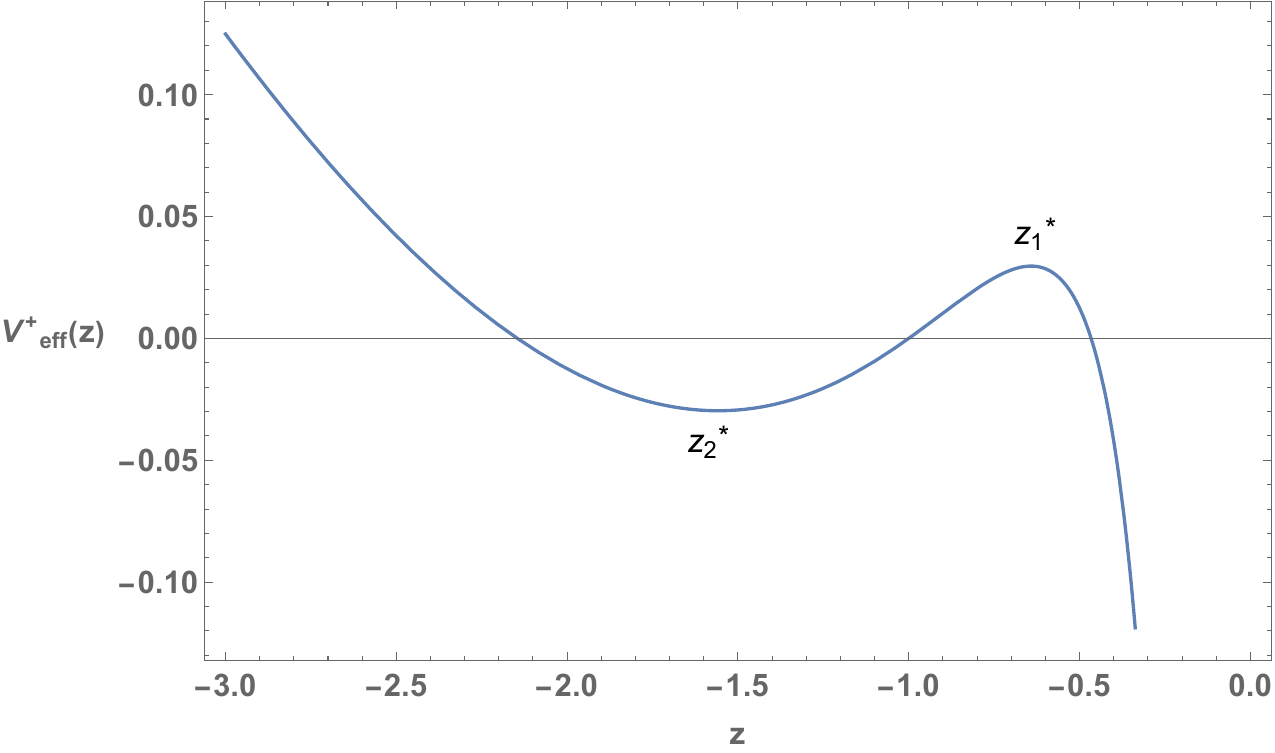}
    \caption{A plot of $\veff^+(z)$ (\ref{veffplus}) on the negative real axis for $t=1.1$. It has a local maximum at $z_1^\star = - t + \sqrt{t^2-1}$ and a local minimum at $z_2^\star = -t-\sqrt{t^2-1}$. The function $\veff^+(z)$ is initially defined outside the unit disk and then analytically continued inside it. The values of $\veff^+$ at the local maximum and minimum are $\pm S_\text{strong}(t)$.}
    \label{fig:plotveffplus}
\end{figure}

The explicit expressions for $\veff^+(z)$ and $\veff^-(z)$ are
\begin{align}
    \veff^{+}(z) &= - \frac{z}{2} + \frac{1}{2z} - t \log (-z) ,  \label{veffplus}\\
    \veff^{-}(z) &= \frac{z}{2} - \frac{1}{2z} + t \log (-z) ,
    \label{veffminus}
\end{align}
A plot of $\veff^+(z)$ for $z<0$ is shown in figure \ref{fig:plotveffplus}.
If we only looked at the original domain of definition of $R^+(z)$, which is outside the unit disk, we might be tempted to compute the action of the instanton as 
\begin{align}
    \veff^+(z_2^\star) = \sqrt{t^2-1}-t \log \left(\sqrt{t^2-1}+t\right) < 0.
\end{align}
Since $\veff^+(z_2^\star) < 0$, this will lead to an \emph{enhanced} contribution $\exp \left( - \frac{N}{t} \veff^+(z_2^\star) \right)$.
The fact the eigenvalue instanton in the strong-coupling phase seems to have the wrong sign was noted earlier in \cite{Buividovich:2015oju}.
However, the defining contour of integration cannot be deformed to the corresponding steepest-descent contour, which is along the real axis (since $\veff^+(z)$ has a local minimum along the real axis at $z_2^\star$).

Now, the main idea is that in order to obtain the instanton contribution in (\ref{main-target}), we should instead consider the contribution proportional to $\exp \left( - \frac{N}{t} \veff^+(z_1^\star) \right)$.
Note that $z_1^\star$ is located inside the unit disk, whereas the original region of definition of $R^+(z)$ in (\ref{rplusdef}) is the exterior of the unit disk.
In this sense, the contribution can be labelled as a ghost-instanton since it represents an eigenvalue tunneling to a location on the ``unphysical sheet'' \cite{Marino:2022rpz}.
We note that
\begin{align}
    \exp \left( - \frac{N}{t} \veff^+(z_1^\star) \right) &= 
    \exp \left( - N \left(\log(t + \sqrt{t^2-1}) - \sqrt{1-t^{-2}} \right) \right)
    = \exp ( - N S_\text{strong}(t))\, .
\end{align}
It is encouraging that the quantity $S_\text{strong}(t)$, defined in (\ref{sstrong}), appears in this expression.

At this stage of the analysis, it is possible that we have contributions of order $\exp ( - N S_\text{strong}(t))$.
It is a well known result from other means of analysis that the leading correction is a two-instanton effect \cite{Marino:2008ya, Ahmed:2017lhl, Okuyama:2017pil}.\footnote{This is true even in the weak-coupling phase \cite{Marino:2008ya}, although for a qualitatively different reason. See also footnote \ref{footweak}.}
We show directly in appendix \ref{oneeigenvalueinstanton} that the contribution to the partition function coming from the tunneling of a single eigenvalue vanishes.
More precisely, if we try to pull out one eigenvalue from the cut, we are forced to pull out a second one as well in order to get a nonzero result.
Thus, the putative one-instanton contribution becomes a two-instanton contribution, and the approach of appendix \ref{oneeigenvalueinstanton} provides an alternative derivation of the result of this section.
However, we note that there are other observables that do receive contributions from one-eigenvalue instantons.
One such example is $\ev{\det U}$ \cite{Green:1981mx, Rossi:1982vw, Okuyama:2017pil}, which we analyze in appendix \ref{appdet} using a direct instanton calculation in the eigenvalue integral.

Thus, let us move on to analyzing two-instanton contributions.
Using the methods of appendix \ref{oneeigenvalueinstanton}, it is possible to show that there is no contribution of order $e^{-2N S_\text{strong}(t)}$ from a configuration where two eigenvalues are placed at the same extremum of the effective potential.
So we consider the configuration where one eigenvalue tunnels to $z_1^\star$ and another tunnels to $z_2^\star$. 
We denote this contribution by $Z^{(1,1)}(N,t)$, which has the following explicit expression:
\begin{multline}
    Z^{(1,1)}(N,t) = \frac{1}{(N-2)!} (-1)^{N(N-1)/2} \int_{\cC_1} \frac{\d z_1}{2\pi\i z_1^N} e^{-\frac{N}{t} V(z_1)} \int_{\cC_2} \frac{\d z_2}{2\pi\i z_2^N} e^{-\frac{N}{t} V(z_2)} (z_1 - z_2)^2 \\
    \times \int_{\cC_0} \prod_{i=3}^{N} \frac{\d z_i}{2\pi\i z_i^N} 
    e^{ -\frac{N}{t} \sum_{i=3}^{N}  V(z_i)} 
    \times \prod_{j=3}^{N} \big[ (z_1 - z_j)^2 (z_2 - z_j)^2 \big] \times \prod_{3\leq k < l \leq N} (z_k - z_l)^2 \, .\label{eqn:Z11start}
\end{multline}
Here, the tunnelled eigenvalues are denoted by $z_1$ and $z_2$ and these variables are integrated along the steepest-descent contours of the respective saddle points. 

We start by considering $\vert z_1 \vert > 1$ and $ \vert z_2 \vert < 1$ and rewrite the above expression as
\begin{multline}
    Z^{(1,1)}(N,t) = \frac{1}{(N-2)!} (-1)^{N(N-1)/2} \int_{\cC_1} \frac{\d z_1}{2\pi\i z_1^N} e^{-\frac{N}{t} V(z_1)} \int_{\cC_2} \frac{\d z_2}{2\pi\i z_2^N} e^{-\frac{N}{t} V(z_2)} (z_1 - z_2)^2 \\
    \times z_1^{2(N-2)} \int_{\cC_0} \prod_{i=3}^{N} \frac{\d z_i}{2\pi\i z_i^{N-2}} e^{ -\frac{N}{t} \sum_{i=3}^{N}  V(z_i)} \times \prod_{j=3}^{N} \big[ (1 - z_j/z_1)^2 (1 - z_2/z_j)^2 \big] \times \prod_{3\leq k < l \leq N} (z_k - z_l)^2 \, .\label{eqn:222}
\end{multline}
This expression will be useful since it contains the factors $(1-z_j/z_1)$ and $(1-z_2/z_j)$ which are such that $\vert z_j / z_1 \vert < 1$ and $\vert z_2/z_j \vert < 1$.
We can now perform the $z_i$ integrals, which is like computing a specific correlator in the GWW matrix integral of rank $N-2$ and with $t \to t - 2t/N$ \cite{Marino:2007te, Marino:2008vx}:
\begin{equation}
\begin{split}
\label{eq:Z11_ev}
    Z^{(1,1)}(N,t) 
    &= - \int_{\cC_1} \frac{\d z_1}{2\pi\i} \int_{\cC_2} \frac{\d z_2}{2\pi\i} e^{-\frac{N}{t} \big[ V(z_1) + V(z_2) \big]} \times \frac{z_1^{N-4} (z_1 - z_2)^2 }{z_2^N} \\
    & \hspace{0.75in} \times Z^{(0)}(N-2,t-2t/N) \ev{ e^{\Tr \log \big[ (1 - z_1^{-1} U )^2 (1 - z_2 U^{-1})^2 \big] } }_{(N-2,t-2t/N)} \\
    &=  - \int_{\cC_1} \frac{\d z_1}{2\pi\i} \int_{\cC_2} \frac{\d z_2}{2\pi\i} e^{-\frac{N}{t} \big[ V(z_1) + V(z_2) \big]} \times \frac{z_1^{N-4} (z_1 - z_2)^2 }{z_2^N} \\
    & \hspace{0.75in} \times Z^{(0)}(N,t) \exp \left[ - \frac{N}{t} \frac{1}{z_1} - \frac{N}{t} z_2 - 4 \log(1 - z_2/z_1) \right] \, .
\end{split}
\end{equation}
Here we have used $Z^{(0)}(N-2,t-2t/N) = e^{N^2/4t^2} = Z^{(0)}(N,t)$.
We have also used the result for the required correlator given in equation \eqref{eq:ev_exp_trlog_2}, the details of which are presented in appendix \ref{apperturbative}.
We can express this result in terms of the effective potentials (\ref{veffplus}) and (\ref{veffminus}) as
\begin{equation}
\label{Z11_Veff}
    \frac{Z^{(1,1)}(N,t)}{Z^{(0)}(N,t)} = - \int_{\cC_1} \frac{\d z_1}{2\pi\i} \int_{\cC_2} \frac{\d z_2}{2\pi\i} e^{-\frac{N}{t} \big[ \veff^{+}(z_1) + \veff^{-}(z_2) \big]} \frac{1}{(z_1 - z_2)^2}.
\end{equation}

Let us now be a bit more explicit about the contours.
The contour $\cC_1$ passes through $z_1^\star$ and $\cC_2$ passes through $z_2^\star$.
Note that $\vert z_1^\star \vert < 1$ and so this lies outside the domain from which we originally started analyzing the $z_1$ integral.
This is the reason why these eigenvalue instantons are referred to as ``ghost instantons.''
A similar comment applies for $z_2$.
The steepest-descent contours at these extrema are along the purely imaginary direction.

The integrals can be evaluated using the saddle-point approximation to get the final answer
\begin{equation}
\begin{split}
    \frac{Z^{(1,1)}(N,t)}{Z^{(0)}(N,t)} 
    &= - \frac{t}{2\pi N} \frac{1}{|\partial^2 \veff^{+}(z_1^\star)|^{1/2} |\partial^2 \veff^{-}(z_2^\star)|^{1/2}} \frac{1}{(z_1^\star - z_2^\star)^2} e^{ - 2 NS_{\text{strong}}(t)} \\
    &= - \frac{t}{8 \pi N (t^2 - 1)^{3/2}} e^{ - 2 NS_{\text{strong}}(t)} \, .
\end{split}
\label{eqn:11contribution}
\end{equation}
Thus, we have managed to precisely reproduce the result (\ref{main-target}) for the leading instanton correction to the GWW partition function in the strong-coupling phase.

The analysis in this section derived the leading nonperturbative contribution to the GWW integral by tunneling eigenvalues to the unphysical sheet of the spectral curve. 
In \cite{Marino:2022rpz}, it was realized that, at least in Hermitian matrix integrals, such ghost instantons can be equivalently realized by tunneling ``negative-charge eigenvalues'' to the corresponding locations on the \emph{physical} sheet.
This leads to a very interesting description of ghost instantons in terms of saddles of an associated supermatrix integral. 
We explore this description for the GWW matrix integral in appendix \ref{supermatrixapp} and reproduce equation (\ref{eqn:11contribution}) from this perspective.

\section{The Fredholm determinant expansion of the GWW integral}
\label{sec:fredholm}
The GWW integral (\ref{defznt}) equals the Toeplitz determinant (\ref{toeplitz}) \cite{Goldschmidt:1979hq}. 
The so-called Fredholm determinant expansion of such Toeplitz determinants is given by the convergent expansion \cite{borodin2000fredholm, geronimo1979scattering}
\begin{equation}
    \frac{Z_N(\tau)}{Z_\infty(\tau)} 
    = \sum_{m=0}^{\infty} (-1)^{m} \sum_{\stac{\{ r_i \} : N \leq r_1 < \cdots r_m }{r_i \in \mathbb{Z} + 1/2}}
    \det \begin{pmatrix} 
    \widetilde{K}(r_1,r_1;\tau) & \widetilde{K}(r_1,r_2;\tau) & \dots & \widetilde{K}(r_1,r_m;\tau) \\
    \vdots & \vdots & \ddots & \vdots \\
    \widetilde{K}(r_m,r_1;\tau) & \widetilde{K}(r_m,r_2;\tau) & \cdots & \widetilde{K}(r_m,r_m;\tau)
    \end{pmatrix} \, .
\label{detexp}
\end{equation}
The quantity $\widetilde{K}(r,s;\tau)$ is given by the generating function
\begin{align}
    \sum_{r,s \in \mathbb{Z} + \frac{1}{2}} 
    \widetilde{K}(r,s;\tau) z^r \, w^{-s} := 
    \frac{\mathcal{J}(z;\tau)}{\mathcal{J}(w;\tau)} 
    \frac{\sqrt{zw}}{z-w} \, , 
    \quad \text{with } \vert w \vert < \vert z \vert \, ,
    \label{defktilde}
\end{align}
and with $\mathcal{J}(z;\tau)$ defined as\footnote{We have specialized the definition of $\mathcal{J}(z;\tau)$ to the GWW integral. The more general expression can be found in \cite{borodin2000fredholm, geronimo1979scattering, Murthy:2022ien}.}
\begin{align}
    \mathcal{J}(z;\tau) := \exp \left(\tau(z - z^{-1}) \right)
    = \sum_{n \in \mathbb{Z}} z^n J_n(2\tau)\, .
\end{align}
Here $J_n$ denotes the Bessel-J function, and the last equality is the well known generating function of the Bessel-J functions.

This Fredholm determinant expansion was used in \cite{Murthy:2022ien} to give a convergent ``giant graviton'' expansion for unitary matrix integrals that arise in the computation of supersymmetric indices in gauge theories.
Since giant gravitons are related to wrapped D-branes, and D-instantons are related to eigenvalue instantons, this suggests that the giant-graviton corrections to the index are related to eigenvalue instantons.\footnote{We thank Steve Shenker for suggesting this connection.}
The unitary matrix integrals arising in the computation of the supersymmetric index differ from the GWW model in that they contain double-trace terms in the action (of the form $\Tr U \Tr U^{-1}$).
While we have not succeeded in finding an eigenvalue instanton interpretation for the giant graviton expansion of these matrix integrals, we show in this section that for the GWW model, which has a single trace action, the first Fredholm-determinant correction in the 't Hooft limit is precisely equal to the leading instanton correction to the partition function (\ref{main-target}). 

As mentioned in the introduction, we always work in the strongly-coupled phase of the GWW integral, $t>1$.
This is because the denominator on the left side of (\ref{detexp}) refers to a limit where $\tau$ (and not $t$) is held fixed as $N \to \infty$.
Recall from equation (\ref{auxdef}) that $\tau = \frac{N}{2t}$.
At this stage of the analysis, it is possible that (\ref{detexp}) is not relevant for the 't Hooft scaled model (one would need to take $t$ of order $N$), but we will see that the $m=1$ term agrees with the first eigenvalue-instanton contribution in the 't Hooft limit for all $t>1$.

We can simplify the ratio of $\mathcal{J}$'s appearing on the right side of (\ref{defktilde}) as
\begin{align}
    \frac{\mathcal{J}(z;\tau)}{\mathcal{J}(w;\tau)}
    &= \exp (\tau(z-z^{-1}))\exp (\tau(w^{-1}-w)) \\
    &= \sum_{n,m \in \mathbb{Z}} z^n w^{-m} 
    J_n(2\tau) J_m(2\tau)\, , 
\end{align}
and  get
\begin{align}
    \frac{\mathcal{J}(z;\tau)}{\mathcal{J}(w;\tau)}
    \frac{\sqrt{zw}}{z-w}
    &= \sum_{n,m \in \mathbb{Z}} z^{n-\frac{1}{2}} w^{-m+\frac{1}{2}}
    J_n(2\tau) J_m(2\tau) \left(1 + \frac{w}{z} + \frac{w^2}{z^2} + \ldots \right)\, .
\end{align}
Comparing the left side of \eqref{defktilde} and the right side of this last equation for $r=s$, we find
\begin{align}
    \widetilde{K}(r,r;\tau) = \sum_{n = r + \frac{1}{2}}^\infty 
    J_n(2\tau)^2\, , \quad r \in \mathbb{Z} + \frac{1}{2}\, .
\end{align}

We will only be interested in the $m=1$ term in \eqref{detexp}, which becomes
\begin{align}
    \frac{Z_N(\tau)}{Z_\infty(\tau)} &= 1 - 
    \sum_{r = N+\frac{1}{2}}^\infty \widetilde{K}(r,r;\tau) 
    + (\text{terms with $m \geq 2$}) 
    \\
    &= 1 - \sum_{k=1}^\infty k\, J_{N+k}(2\tau)^2
    + (\text{terms with $m \geq 2$}) \, .
    \label{gn1result}
\end{align}
Note that the index of the Bessel functions appearing in the final expression is large, greater than $N$.
We would like this expression to reproduce \eqref{main-target}, including the action of the instanton and the one-loop determinant.

In order to do so, we need to work in the 't Hooft limit, and thus consider $\tau$ of order $N$.
In this regime, both the index and the argument of the Bessel-J functions appearing in \eqref{gn1result} are large.
Therefore, Debye's expansion of the Bessel-J function \cite{debye} is valid:
\begin{align}
    J_N\left( \frac{N}{\cosh \alpha}\right) = 
    e^{-N(\alpha - \tanh \alpha)} \, (2\pi N \tanh \alpha)^{-\frac{1}{2}} \left(1 + O\left(\frac{1}{N}\right) \right) \, .
    \label{debyeJ}
\end{align}
The right side of this expression looks exactly like a saddle-point contribution.
Comparing this to \eqref{debyeJ} is encouraging.
The on-shell action seems to be just right, after setting $\cosh \alpha = t$:
\begin{align}
    \alpha  - \tanh \alpha = \arccosh(t) - \sqrt{1-t^{-2}} = S_\text{strong}(t) \, .
\end{align}
However, we need to do some more work to get the correct one-loop factor.
We will get this from the infinite sum over the integer $k$ in \eqref{gn1result}, and the results will match.

Let us delve into the details of the one-loop prefactor.
Define $\alpha_k$ and $\beta_k$ using
\begin{align}
    \frac{N}{t} = \frac{N+k}{\cosh \alpha_k} = (N+k) \, \beta_k\, .
\end{align}
Now using Debye's expansion, we get
\begin{equation}
    J_{N+k}\left(\frac{N}{t} \right)^2 \approx \frac{1}{2\pi(N+k) \sqrt{1-\beta_k^2}} \left( \frac{\beta_k \, e^{\sqrt{1-\beta_k^2}}}{1 + \sqrt{1-\beta_k^2} } \right)^{2N+2k}\, .
    \label{debye2}
\end{equation}
When summing over $k$, we need to be careful about taking
\begin{equation}
    \beta_k = \frac{N/t}{N+k} \approx \frac{1}{t} \left(1 - \frac{k}{N} \right).
    \label{betakapprox}
\end{equation}
Since the right side of \eqref{debye2} involves a large exponent, corrections of order $1/N$ inside the parentheses can yield order-one results.
Note that the approximation \eqref{betakapprox} only makes sense for $k \ll N$, but we will use it to do the sum in \eqref{gn1result}.
The corrections to this approximation become very important at $k \sim N$, but then the corresponding terms become comparable to the contribution from the second nontrivial term in the Fredholm determinant expansion, and we are not working to that accuracy.

Being careful about the $1/N$ corrections to $\beta_k$ discussed above, we obtain
\begin{align}
    \frac{J_{N+k}(N/t)^2}{J_{N}(N/t)^2} 
    &\approx \left( \frac{1}{t + \sqrt{t^2-1} } \right)^{2 k}.
\end{align}
Noting also from (\ref{debyeJ}) that
\begin{align}
    J_{N} \left(N/t\right)^2 
    &\approx e^{-2 N S_\text{strong}(t)} \cdot \frac{1}{2\pi N \sqrt{1-1/t^2}} \, ,
\end{align}
we compute the contribution of one giant graviton (\ref{gn1result}) as
\begin{align}
    \frac{Z(N,t)}{e^{N^2/4t^2}} - 1 
    &= - \sum_{k=1}^{\infty} k \, J_{N+k}(N/t)^2 \\
    &\approx - e^{-2 N S_\text{strong}(t)}\cdot \frac{1}{2\pi N \sqrt{1-1/t^2}} \sum_{k=1}^{\infty} k \left( \frac{1}{t + \sqrt{t^2-1} } \right)^{2 k}  \\
    &= -  e^{-2 N S_\text{strong}(t)} \,\frac{1}{N} \, \frac{t}{8 \pi (t^2-1)^{3/2} } \,.
    \label{match}
\end{align}
The result \eqref{match} agrees precisely with our target \eqref{main-target}.

\section{Summary and Discussion}
\label{sec:summary}

In this paper we have analyzed the leading nonperturbative correction to the Gross-Witten-Wadia integral (\ref{defznt}) in the strong-coupling phase, $t>1$.
In the past, the result (\ref{main-target}) was derived using techniques such as orthogonal polynomials and by studying differential equations satisfied by various observables \cite{Marino:2008ya, Ahmed:2017lhl, Okuyama:2017pil}.

In section \ref{sec:eiginst}, we showed that it is possible to derive (\ref{main-target}), which is a two-eigenvalue instanton effect, by expressing the GWW integral as a holomorphic integral over its eigenvalues (\ref{zntholomorphic1}) and directly analyzing the instanton effects in this integral.
The integral expression for the two-eigenvalue instanton contribution involving the effective potentials is given in (\ref{Z11_Veff}), correct to one-loop order.
We also showed in appendix \ref{oneeigenvalueinstanton} that there are no one-eigenvalue instanton contributions to the partition function, and in appendix \ref{appdet} that there are other observables such as $\ev{\det U}$ that do receive contributions from one-eigenvalue instantons.
All these computations were done via a direct analysis of the integrals (\ref{defznt}) and (\ref{zntholomorphic1}), and involve eigenvalue tunneling to complex locations on the unphysical sheet.
Such instantons are called ghost instantons and their action has the opposite sign compared to their partners on the physical sheet.

The GWW integral admits a double-scaling limit, and in this limit, it is dual to the minimal superstring \cite{Klebanov:2003wg}.
An interesting future goal is to reproduce the precise nonperturbative corrections from ZZ-instantons in the minimal superstring, generalizing previous work in the $c=1$ bosonic string \cite{Balthazar:2019rnh, Sen:2021qdk}, the $c<1$ bosonic string \cite{Eniceicu:2022nay, Eniceicu:2022dru} and the $\hat{c}=1$ type 0B string \cite{Chakravarty:2022cgj}. 
However, we note that the results of the present paper predict a very surprising organization of the asymptotics of the minimal superstring partition function. 
Usually, in perturbative string theory, the partition function is described by the exponential of a perturbative genus expansion associated to closed surfaces, potentially supplemented by nonperturbative D-brane effects. 
In contrast to the usual case, in the minimal superstring dual to the strong-coupling phase of the GWW model, the perturbative contribution (sum over closed surfaces) to the partition function is entirely absent. 
The leading asymptotics are instead associated to the exponential of disk and cylinder diagrams for a pair of ghost ZZ branes. 
We are currently investigating this problem \cite{wip}.

In section \ref{sec:fredholm}, we showed that the first nontrivial term in the Fredholm determinant expansion of the Toeplitz determinant (\ref{toeplitz}), in the 't Hooft limit and to one-loop order, precisely reduces to the result (\ref{main-target}).
We only analyzed the strong-coupling phase of the GWW model ($t>1$), but the weak-coupling phase ($t<1$) also has eigenvalue instanton effects \cite{Marino:2008ya}.
An open problem is whether these can also be related to the Fredholm determinant expansion.

Our motivation for studying this connection was to provide an eigenvalue instanton interpretation for the giant-graviton expansion studied in the context of supersymmetric index computations in \cite{Murthy:2022ien}.
The unitary matrix integrals in \cite{Murthy:2022ien} have double trace terms in the action, and these are related to matrix integrals with a single-trace action (like the GWW model) via a Hubbard-Stratonovich transform.
We leave it for future work to study the fate of the eigenvalue instantons after doing a Hubbard-Stratonovich transform.
However, we are tempted to make some preliminary remarks about this problem.

For this, we consider a simple integral with the action containing a single double-trace term \cite{Copetti:2020dil}.
\begin{align}
    Z(N,a_1) := \int \frac{\d U}{\vol U(N)} \, \exp \left( 
    a_1 \Tr U \Tr U^{-1}
    \right)\, .
\end{align}
This model has a phase transition at $a_1 = 1$.
We work in the phase with $a_1 < 1$ for which the eigenvalue density is ungapped i.e., supported on the entire unit circle.
We first write this integral as Hubbard-Stratonovich transform of the GWW integral and then use the Fredholm determinant expansion of the GWW integral.
Directly, we can use Theorem 5.2 in \cite{Murthy:2022ien}, specialized to the present case, to conclude that
\begin{align}
    Z(N,a_1) = \frac{1}{1-a_1} \left( 1 - \frac{a_1^{N+1}}{(N+1)!} + O(a_1^{N+2})\right)\, .
\end{align}
Note that the leading correction is of order $e^{- N \log N}$, which is parametrically smaller than what is expected from eigenvalue instantons (which should be of the form $e^{-c N}$). 

Now, the unitary matrix integrals that show up in the computation of supersymmetric indices have an infinite number of double-trace couplings turned on.
It is possible that models with an infinite number of terms behave qualitatively differently than this simple ``$a_1$-model.''
We leave it for future work to explore this further.

\paragraph{Acknowledgments.}
We would like to thank Abhijit Gadde, Ohad Mamroud, Sameer Murthy, Steve Shenker, Douglas Stanford, and Mykhaylo Usatyuk for discussions. 
D.S.E.~is supported by the Shoucheng Zhang Graduate Fellowship.
C.M.~is supported in part by the U.S.~Department of Energy, Office of Science, Office of High Energy Physics under QuantISED Award DE-SC0019380 and contract DE-AC02-05CH11231.

\appendix

\section{Further details about eigenvalue instanton computations}
\label{app}

\subsection{Some perturbative correlators}
\label{apperturbative}

In this subsection, we explain the computation of $\ev{ e^{\Tr \log \big[ (1 - z_1^{-1} U )^2 (1 - z_2 U^{-1})^2 \big] } }_{(N,t)}$ that is needed in (\ref{eq:Z11_ev}).
From now on, we will suppress the $(N,t)$ subscript on the expectation value.
Also recall that we would like to define this correlator initially for $\vert z_1 \vert > 1$ and $\vert z_2 \vert < 1$, and then analytically continue the expression obtained.

Using the identity 
$\ev{e^{X}} = \exp \left[ \ev{X} + \frac{1}{2} \ev{X X}_c + \ldots \right]$, we have
\begin{multline}
\label{eq:ev_exp_trlog_1}
    \ev{ e^{\Tr \log \big[ (1 - z_1^{-1} U )^2 (1 - z_2 U^{-1})^2 \big] } }
    \approx \exp \Big[ 2 \ev{\Tr \log (1 - z_1^{-1} U )} + 2 \ev{\Tr \log (1 - z_2 U^{-1} )} \\
    + 2 \ev{\Tr \log (1 - z_1^{-1} U ) \Tr \log (1 - z_1^{-1} U )}_c + 2 \ev{\Tr \log (1 - z_2 U^{-1} ) \Tr \log (1 - z_2 U^{-1} )}_c \\
    + 4 \ev{\Tr \log (1 - z_1^{-1} U ) \Tr \log (1 - z_2 U^{-1} )}_c\Big] \, .
\end{multline}
Let us now evaluate the various correlators present on the right side. 
The starting point for these computations is the single-trace correlator
\begin{equation}
    \ev{\Tr U^j} = 
    \begin{cases}
        N \, , & j = 0 \, , \\
        \frac{N}{2t} \, , & j = \pm 1 \, , \\
        0 \, , & \text{otherwise,}
    \end{cases}
    \label{singletrace-uj}
\end{equation}
and the double-trace connected correlator
\begin{equation}
    \ev{\Tr U^j \Tr U^k}_c = |j| \, \delta_{j,-k} \, .
    \label{double-trace-ujuk}
\end{equation}
These are evaluated by expanding the exponential in the GWW measure (\ref{defznt}) as in the computation of $Z^{(0)}(N,t)$ in (\ref{zntperturbative}), and using the result (\ref{diacshah}).

The result (\ref{singletrace-uj}) allows us to compute the single-trace correlators appearing in (\ref{eq:ev_exp_trlog_1}) by doing a Taylor series expansion
\begin{align}
    \ev{\Tr \log (1 - z U^{-1}) }
    &= - \sum_{j=1}^{\infty} \frac{1}{j} z^j \ev{\Tr U^{-j}} = - \frac{N}{2t} z \, , 
    & |z| < 1 \, , \label{trlog1} \\
    \ev{\Tr \log (1 - z^{-1} U) }
    &=  - \sum_{j=1}^{\infty} \frac{1}{j} z^{-j} \ev{\Tr U^{j}} = - \frac{N}{2t} \frac{1}{z} \, , 
    & |z| > 1 \, . \label{trlog2}
\end{align}
Using the result (\ref{double-trace-ujuk}), we see that two of the double-trace connected correlators appearing in (\ref{eq:ev_exp_trlog_1}) vanish:
\begin{align}
    \ev{\Tr \log (1 - z_1 U^{-1}) \Tr \log (1 - z_2 U^{-1}) }_c
    &= \sum_{j,k=1}^{\infty} \frac{1}{jk} z_1^j z_2^k \ev{\Tr U^{-j} \Tr U^{-k} }_c =  0 \, ,
    & |z_1|,|z_2|<1 \, ,  \label{trlog3}\\
    \ev{\Tr \log (1 - z_1^{-1} U) \Tr \log (1 - z_2^{-1} U) }_c
    &= \sum_{j,k=1}^{\infty} \frac{1}{jk} z_1^{-j} z_2^{-k} \ev{\Tr U^{j} \Tr U^{k} }_c =  0 \, ,
    & |z_1|,|z_2| > 1 \, , \label{trlog4}
\end{align}
while the third one is
\begin{align}
    \ev{\Tr \log (1 - z_1^{-1} U) \Tr \log (1 - z_2 U^{-1}) }_c
    &= \sum_{j,k=1}^{\infty} \frac{1}{jk} z_2^{j} z_1^{-k} \ev{\Tr U^{-j} \Tr U^{k} }_c 
    \nonumber \\
    &= - \log ( 1 - z_2/z_1) \, , 
    \quad\quad\quad
    |z_1|>1 \, ,|z_2| < 1 \, . \label{trlog5}
\end{align}
Substituting the results (\ref{trlog1})-(\ref{trlog5}) into (\ref{eq:ev_exp_trlog_1}), we get the final result
\begin{equation}
    \label{eq:ev_exp_trlog_2}
    \ev{ e^{\Tr \log \big[ (1 - z_1^{-1} U )^2 (1 - z_2 U^{-1})^2 \big] } }_{(N,t)} \approx \exp \left[ - \frac{N}{t} \frac{1}{z_1} - \frac{N}{t} z_2 - 4 \log(1 - z_2/z_1) \right] \, .
\end{equation}
This is the correlator that we needed for the computation of $Z^{(1,1)}(N,t)$ in (\ref{eq:Z11_ev}).

\subsection{Expectation value of $\det U$}
\label{appdet}

Even though there is no one-instanton contribution to the partition function, there are other observables that do receive contributions from single instantons.
One such quantity is $\ev{\det U}$, which we analyze in this appendix.
This was also discussed in \cite{Okuyama:2017pil} from the perspective of orthogonal polynomials. 
Our goal is to redo the computation directly in terms of a one-eigenvalue instanton in the integral (\ref{zntholomorphic1}).

It is known that the quantity $\ev{\det U}$ vanishes to all orders perturbation theory in the strong-coupling phase of the GWW model \cite{Ahmed:2017lhl, Rossi:1982vw, Green:1981mx}.
As usual, one way to see this is to series expand the exponential in the defining integral (\ref{defznt}) and then expand $(\Tr U + \Tr U^{-1})^n$ via the binomial theorem. 
A generic term will look like 
\begin{equation}
    \left(\frac{N}{2t}\right)^n \frac{1}{k!(n-k)!} \int \frac{\d U}{\vol U(N)} \, \det U (\Tr U)^k (\Tr U^{-1})^{n-k} \, ,
\end{equation}
with $0 \leq k \leq n$.
Making a change of variables $U \to e^{\i \alpha} U$, we see that we must have $N = n - 2k$ in order to get a nonzero result.
In particular, this means that $n \geq N$. 
The first nonzero term in the series (with $n=N, k=0$) is thus exponentially small in $N$.
See also the discussion below (\ref{detzrelation}).

In this subsection, we show that the leading contribution to $\ev{\det U}$ in the strong-coupling phase comes from a one-eigenvalue instanton.
Pulling out one eigenvalue from the cut, and placing it at some position $z_1$ outside the unit circle, we have the expression
\begin{align}
    \ev{\det U}_{(N,t)}^{(1)}
    &= \frac{1}{Z^{(0)}(N,t)} \times \frac{1}{(N-1)!} (-1)^{N(N-1)/2} \int_{\cC_1} \frac{\d z_1}{2\pi\i z_1^N} e^{-\frac{N}{t} V(z_1)} \nonumber \\
    & \hspace{0.5in} \times \int_{\cC_0} \prod_{i=2}^{N} \frac{\d z_i}{2\pi\i z_i^N} e^{-\frac{N}{t} \sum_{i=2}^{N}  V(z_i)} \times \prod_{j=2}^{N}  (z_1 - z_j)^2 \times \prod_{2\leq j < k \leq N} (z_j - z_k)^2 \times \prod_{j=1}^{N} z_j  \nonumber\\
    &= \frac{1}{Z^{(0)}(N,t)} \times \frac{1}{(N-1)!} (-1)^{N(N-1)/2} \int_{\cC_1} \frac{\d z_1}{2\pi\i z_1^{N-1}} e^{-\frac{N}{t} V(z_1)}  \nonumber \\
    & \hspace{0.5in} \times \int_{\cC_0} \prod_{i=2}^{N} \frac{\d z_i}{2\pi\i z_i^{N-1}} e^{- \frac{N}{t} \sum_{i=2}^{N}  V(z_i)} \times \prod_{j=2}^{N} \big[ z_1^2 (1 - z_j/z_1)^2 \big] \times \prod_{2\leq j < k \leq N} (z_j - z_k)^2 \nonumber \\
    &= (-1)^{N-1} \int_{\cC_1} \frac{\d z_1}{2\pi\i z_1^{N-1}} e^{-\frac{N}{t} V(z_1)} \times z_1^{2(N-1)}\times \ev{e^{2 \Tr \log(1- z_1^{-1} U)}}_{(N-1,t-\frac{t}{N})}  \nonumber \\
    &\approx (-1)^{N-1} \int_{\cC_1} \frac{\d z_1}{2\pi\i } e^{-\frac{N}{t} V(z_1)} \times z_1^{(N-1)} \times e^{- \frac{N}{t} \frac{1}{z_1} } \nonumber \\
    &= - \int_{\cC_1} \frac{\d z_1}{2\pi\i z_1} e^{-\frac{N}{t} \veff^{+}(z_1)} .
\end{align}
We used the fact that $Z^{(0)}\left(N-1,t-\frac{t}{N}\right)  = Z^{(0)}\left(N,t\right)$,
We have used the results (\ref{trlog2}) and (\ref{trlog4}) to get the expression in the second-to-last line.
In the last line, we have used (\ref{veffplus}).

As in the case of the partition function itself, even though we started by considering $z_1$ to be outside the unit circle, the relevant saddle point lies inside the unit circle.
The defining integration contour is such that we need to pick the orientation of the steepest-descent contour as being vertically \emph{upwards} through $z_1^\star$.
Doing a saddle point evaluation at $z_1 = z_1^\star$, we get
\begin{equation}
    \ev{\det U}_{(N,t)}^{(1)} 
    \approx \frac{1}{2\pi (-z_1^\star)} e^{-\frac{N}{t} \veff^{+}(z_1^\star)}
    \sqrt{\frac{2\pi t}{N |\partial^2 \veff^{+}(z_1^\star)| }} 
    = e^{-N S_{\text{strong}}(t)} \frac{1}{\sqrt{N}} \, 
    \frac{\sqrt{t/2\pi}}{(t^2-1)^{1/4}}.
    \label{detU}
\end{equation}
This is indeed the known leading contribution to $\ev{\det{U}}$ in the strong-coupling phase of the GWW model \cite{Ahmed:2017lhl}.

We can do a quick consistency check of this result, which we have derived from saddle-point analysis of the integral over eigenvalues.
There is a known relationship between $\ev{\det U}_{N,t}$ and the partition function \cite{Marino:2008ya, Ahmed:2017lhl}
\begin{align}
    \ev{\det U}_{(N,t)}^2 = 1 - \frac{Z(N+1,t+\frac{t}{N}) \, Z(N-1,t-\frac{t}{N})}{Z(N,t)^2}\, .
    \label{detzrelation}
\end{align}
First note that we can use (\ref{detzrelation}) and the result $Z^{(0)}(N,t) = \exp(N^2/4t^2)$ to show that $\ev{\det U}$ vanishes to all orders in perturbation theory (when $t>1$).
Second, it is straightforward to plug in (\ref{detU}) on the left side and (\ref{main-target}) on the right side and verify the equality (\ref{detzrelation}).

\subsection{The putative one-eigenvalue instanton contribution to $Z(N,t)$}
\label{oneeigenvalueinstanton}

In this subsection, we show that there are no one-eigenvalue instanton contributions (of order $e^{-N S_\text{strong}(t)}$) to $Z(N,t)$ in the strong-coupling phase. 
If we try to pull out one eigenvalue from the cut, we shall see that we are forced to pull out a second one as well in order to get a nonzero result.

This result is known from previous analyses \cite{Marino:2008ya, Okuyama:2017pil, Ahmed:2017lhl} via other methods.
In fact, it is also true in the weak-coupling phase that the leading instanton correction to the perturbation series for the partition function is a two-instanton effect \cite{Marino:2008ya}.\footnote{In the weak-coupling phase, the eigenvalue instanton is a maximum of the effective potential that lies on the unit circle, which is the original integration contour. There is a general argument that such saddle points cannot contribute to the integral. A quick way to see this is that it would yield a purely imaginary contribution to the partition function, which is ruled out since the partition function is purely real. In an analysis using the Borel transform, this is attributed to median resummation, or the cancellation of nonperturbative ambiguities \cite{Marino:2008ya}.\label{footweak}}
Our goal is to see this explicitly via an analysis of the eigenvalue instantons in the integral (\ref{zntholomorphic1}).

Pulling out just one eigenvalue $z_1$ and placing it outside the unit circle, the putative one-eigenvalue instanton contribution to $Z(N,t)$ is given by
\begin{multline}
    Z^{(1)}(N,t) = \frac{1}{(N-1)!} (-1)^{N(N-1)/2} \int_{\cC_1} \frac{\d z_1}{2\pi\i z_1^N} e^{-\frac{N}{t} V(z_1)}  \\
    \times \int_{\cC_0} \prod_{i=2}^{N} \frac{\d z_i}{2\pi\i z_i^N} 
    e^{-\frac{N}{t} \sum_{i=2}^{N}  V(z_i)} \times \prod_{j=2}^{N} (z_1 - z_j)^2  \times \prod_{2\leq j < k \leq N} (z_j - z_k)^2 .
\end{multline}
We can rewrite the above expression as
\begin{multline}
    Z^{(1)}(N,t) = \frac{1}{(N-1)!} (-1)^{N(N-1)/2} 
    \int_{\cC_1} \frac{\d z_1}{2\pi\i}\, z_1^{N-2} e^{-\frac{N}{t} V(z_1)} \\
    \times \int_{\cC_0} \prod_{i=2}^{N} \frac{\d z_i}{2\pi\i z_i^{N-1}} e^{- \frac{N}{t} \sum_{i=2}^{N}  V(z_i)} \times 
    \prod_{j=2}^{N} \big[ z_j^{-1} (1 - z_j/z_1)^2 \big] 
    \times \prod_{2\leq j < k \leq N} (z_j - z_k)^2 .
\end{multline}
Now we do the $z_2,z_3,\ldots, z_N$ integrals to get
\begin{equation}
    Z^{(1)}(N,t) = (-1)^{N-1} \int_{\cC_1} \frac{\d z_1}{2\pi\i} z_1^{N-2} e^{-\frac{N}{t} V(z_1)} \times Z^{(0)}\left(N-1,t-\frac{t}{N}\right) 
    \ev{ \det U^{-1} \det(1 - z_1^{-1} U)^2 }_{(N-1,t-t/N)} .
\end{equation}
Using the fact that $Z^{(0)}\left(N-1,t-\frac{t}{N}\right)  = Z^{(0)}\left(N,t\right)$, we get
\begin{align}
    \frac{Z^{(1)}(N,t)}{Z^{(0)}(N,t)} = 
    (-1)^{N-1} \int_{\cC_1} \frac{\d z_1}{2\pi\i}z_1^{N-2} e^{-\frac{N}{t} V(z_1)} 
    \ev{ \det U^{-1} \det(1 - z_1^{-1} U)^2 }_{(N-1,t-t/N)} .
    \label{z1overz0}
\end{align}

Similar to $\ev{ \det U }$, the expectation value $\ev{ \det U^{-1} \det(1 - z^{-1} U)^2 }$ vanishes at the perturbative order. 
We can show this by expanding the second term as a power series in $z^{-1}$
\begin{equation}
\begin{split}
    \ev{ \det U^{-1} \det(1 - z^{-1} U)^2 }_{(N,t)} 
    &=  \sum_{k=0}^{\infty} \frac{2^k}{k!} \ev{ \det U^{-1} \left( - \sum_{j=1}^{\infty} \frac{z^{-j}}{j} \Tr U^j \right)^{k} }_{(N,t)} \, .
\end{split}
\end{equation}
Note that the coefficients of $z^{-m}$ vanish for $m>2N$, since $\det (1-z^{-1} U)$ is a polynomial in $z^{-1}$ of degree $N$.
Using an argument similar to the one for $\ev{\det U}$ in appendix \ref{appdet}, we can prove that this quantity is exponentially small in $N$.
Thus, the perturbative contribution to $\ev{ \det U^{-1} \det(1 - z^{-1} U)^2 }$ vanishes, and we now look for eigenvalue instanton contributions to this quantity.

The leading contribution to $\ev{ \det U^{-1} \det(1 - z^{-1} U)^2 }$ comes from a one-eigenvalue instanton, just like $\ev{\det U}$.
Plugging this one-eigenvalue instanton contribution back into (\ref{z1overz0}) will result in an expression identical to (\ref{Z11_Veff}), which is a two-instanton effect. 
Thus, this method yields an alternative derivation of the result (\ref{main-target}).

Pulling out one eigenvalue from the cut, and placing it at some position $z_1$ inside the unit circle, we have the expression
\begin{align}
    &\ev{\det U^{-1} \det(1 - z^{-1} U)^2}^{(1)}_{(N,t)}
    = \frac{1}{Z^{(0)}(N,t)} \times \frac{(-1)^{N(N-1)/2}}{(N-1)!} \int_{\cC_2} \frac{\d z_1}{2\pi\i z_1^{N+1}} e^{-\frac{N}{t} V(z_1)}  (1 - z_1/z)^2 \nonumber \\
    & \hspace{0.2in} \times \int_{\cC_0} \prod_{i=2}^{N} \frac{\d z_i}{2\pi\i z_i^{N-1}} e^{- \frac{N}{t} \sum_{i=2}^{N}  V(z_i)} \times \prod_{j=2}^{N} \big[ (1 - z_1/z_j)^2 (1 - z_j/z)^2 \big]
    \times \prod_{2\leq j < k \leq N} (z_j - z_k)^2 \nonumber \\
    &= (-1)^{N-1} \int_{\cC_2} \frac{\d z_1}{2\pi\i z_1^{N+1}} e^{-\frac{N}{t} V(z_1)} (1 - z_1/z)^2 \times \ev{e^{2 \Tr \log(1- z^{-1} U) + 2 \Tr \log(1- z_1 U^{-1})}}_{(N-1,t-\frac{t}{N})}  \nonumber \\
    &\approx (-1)^{N-1} \int_{\cC_2} \frac{\d z_1}{2\pi\i z_1^{N+1}} e^{-\frac{N}{t} V(z_1)} (1 - z_1/z)^2 \times \frac{ e^{ - \frac{N}{t} \frac{1}{z} - \frac{N}{t} z_1} }{(1 - z_1/z)^4} \nonumber \\
    &= - \int_{\cC_2} \frac{\d z_1}{2\pi\i} e^{-\frac{N}{t} \veff^{-}(z_1)}  e^{- \frac{N}{t} \frac{1}{z}} \frac{1}{z_1 (1 - z_1/z)^2} \, . \label{detdet}
\end{align}
We have used (\ref{trlog1}), (\ref{trlog2}) and (\ref{trlog5}) to get to the second-to-last line.

In (\ref{z1overz0}), we need $\ev{\det U^{-1} \det(1 - z^{-1} U)^2}^{(1)}_{(N-1,t-t/N)}$, with shifted values of $N$ and $t$.
Note that $\veff^{-}(z_1)$ given in (\ref{veffminus}) has an explicit dependence on $t$.
The required shift $t \to t - t/N$ nicely absorbs the $1/z_1$ factor present in the integrand in (\ref{detdet}), so we get
\begin{equation}
    \ev{\det U^{-1} \det(1 - z^{-1} U)^2}_{(N-1,t-t/N)}^{(1)} = \int_{\cC_2} \frac{\d z_2}{2\pi\i} e^{-\frac{N}{t} \veff^{-}(z_2)}  e^{- \frac{N}{t} \frac{1}{z}} \frac{1}{(1 - z_2/z)^2} \, ,
\end{equation}
where we have relabeled the integration variable as $z_2$.
Substituting this into (\ref{z1overz0}), we get 
\begin{align}
    Z^{(1,1)}(N,t) 
    &= (-1)^{N-1} \int_{\cC_1} \frac{\d z_1}{2\pi\i}z_1^{N-2} e^{-\frac{N}{t} V(z_1)} \int_{\cC_2} \frac{\d z_2}{2\pi\i} e^{-\frac{N}{t} \veff^{-}(z_2)}  e^{- \frac{N}{t} \frac{1}{z_1}} \frac{1}{(1 - z_2/z_1)^2} \nonumber \\
    &= - \int_{\cC_1} \frac{\d z_1}{2\pi\i} \int_{\cC_2} \frac{\d z_2}{2\pi\i}  e^{-\frac{N}{t} \veff^{+}(z_1) -\frac{N}{t} \veff^{-}(z_2)}  \frac{1}{(z_1 - z_2)^2}\, .
\end{align}
This is now a two-eigenvalue instanton contribution, which we have indicated by changing $Z^{(1)}(N,t)$ to $Z^{(1,1)}(N,t)$.
This equation is exactly (\ref{Z11_Veff}) and thus gives us an alternative derivation of (\ref{main-target}), the leading nonperturbative correction to $Z(N,t)$ in the strong-coupling phase.

\section{Ghost-instanton contribution from tunneling anti-eigenvalues}
\label{supermatrixapp}
A key insight of \cite{Marino:2022rpz} is the fact that the ghost-instanton contributions to a matrix integral can be described in two equivalent ways: either as the tunneling of regular eigenvalues to the unphysical sheet of the spectral curve, or by upgrading the matrix integral to a supermatrix integral,\footnote{We will exclusively use the ``physical supermatrix integral.'' 
The physical supermatrix integral is defined explicitly in terms of an integral over the eigenvalues of the supermatrix, and should be contrasted with the ``ordinary supermatrix integral,'' which plays no role here. 
See \cite{Marino:2022rpz} and \cite{Kimura:2023iup} for details and equation (\ref{eqn:defsupermatrix}) for the precise definition we will use.} and tunneling a corresponding number of negative-charge eigenvalues to the physical sheet. 
The equivalence between these two pictures is summarized by equations (3.13) and (3.21) of \cite{Marino:2022rpz} for the single-anti-eigenvalue contributions and for the multiple-anti-eigenvalue contributions, respectively.

The goal of this appendix is to reproduce the leading ghost-instanton contribution to the GWW matrix integral in the strong-coupling phase in the supermatrix picture. This complements the derivation in section \ref{sec:eiginst} of the same result, equation (\ref{eqn:11contribution}), which was obtained by tunneling regular eigenvalues to the opposite sheet. Since we will be interested in the contribution from one negative-charge-eigenvalue tunneling to $z_1^\star$ and one negative-charge-eigenvalue tunneling to $z_2^\star$, the starting point will be a unitary $(N+2|2)\times(N+2|2)$ supermatrix integral, defined in terms of the integral over its eigenvalues\footnote{In the following, the notation $\bar{z}$ is used to denote a negative-charge-eigenvalue of the supermatrix. It does not denote complex conjugation.} as:
\begin{align}
    \label{eqn:defsupermatrix}
    Z \left( N+2|2\right) 
    &:= \frac{1}{(N+2)!\;2!} \text{PV}\int_{\cC_0}\prod_{i=1}^{N+2}\frac{\d z_i}{2\pi \i z_i}\int_{\bar{\cC}_0}
    \prod_{j=1}^2\frac{\d\bar{z}_j}{2\pi \i\bar{z}_j}\;\Delta_{N+2|2}(z,\bar{z})^2\times \nonumber \\
    & \hspace{1.5in} \times\exp\left(- \frac{N}{t} \sum_{i=1}^{N+2}V\left(z_i\right)+\frac{N}{t} \sum_{j=1}^2 V\left(\bar{z}_j\right)\right),
\end{align}
where we define the super-Vandermonde determinant
\begin{equation}
    \Delta_{N+2|2}(z,\bar{z}) = \frac{\left|\bar{z}_1-\bar{z}_2\right| \prod_{1\leq i<j\leq N+2} \left|z_i-z_j\right|}{\prod_{i=1}^{N+2}\left|\bar{z}_1-z_i\right| \left|\bar{z}_2-z_i\right|} \, ,
\end{equation}
and the $\text{PV}$ denotes a particular regularization prescription required to render the integral finite \cite{Marino:2022rpz}. 
In the Hermitian case, this prescription corresponds to the Cauchy principal value (see \cite{Marino:2022rpz} for details). 
In the unitary case, a definition was provided in \cite{Eniceicu:2023uvd}.
Since we are solely interested in using the supermatrix integral description to obtain nonperturbative corrections to an ordinary matrix integral, the choice of prescription does not play a role, so we will not attempt to specify it in the present unitary case. 
We can use the identity
\begin{equation}
    \vert z_j - z_k \vert^2 = - \frac{1}{z_j z_k} (z_j - z_k)^2 \, ,
\end{equation}
which is valid on the unit circle, to cast the supermatrix integral in a holomorphic form
\begin{multline}
    Z \left( N+2|2 \right)
    = - \frac{(-1)^{(N+2)(N+1)/2}}{(N+2)!\;2!} \text{PV}\int_{\cC_0}\prod_{i=1}^{N+2}\frac{\d z_i}{2\pi \i z_i^N}\int_{\bar{\cC}_0}\prod_{j=1}^2\frac{\bar{z}_j^N\d\bar{z}_j}{2\pi \i}\times \\
    \times \frac{\left(\bar{z}_1-\bar{z}_2\right)^2\prod_{1\leq i<j\leq N+2}\left(z_i-z_j\right)^2}{\prod_{i=1}^{N+2}\left(\bar{z}_1-z_i\right)^2 \left(\bar{z}_2-z_i\right)^2}
    \times\exp\left(-\frac{N}{t} \sum_{i=1}^{N+2} V \left(z_i\right) + \frac{N}{t} \sum_{j=1}^2V\left(\bar{z}_j\right) \right) \, .
\end{multline}
The leading ghost-instanton correction to the $N\times N$ GWW matrix integral will be obtained from the above expression after modifying the contours of integration for the two anti-eigenvalues to pass through $z_1^\star$ and $z_2^\star$, respectively. 
Since there are two ways to choose which anti-eigenvalue tunnels to which contour, amounting to an overall factor of two, the leading contribution is
\begin{multline}
    Z^{(1,1)}(N,t)=-\frac{(-1)^{(N+2)(N+1)/2}}{(N+2)!} \text{PV}\int_{\bar{\cC}_1^\star}\frac{\bar{z}_1^N\d\bar{z}_1}{2\pi \i}\int_{\bar{\cC}_2^\star}\frac{\bar{z}_2^N\d\bar{z}_2}{2\pi \i}\left(\bar{z}_1-\bar{z}_2\right)^2\exp\left(\frac{N}{t}\sum_{j=1}^2V\left(\bar{z}_j\right)\right)\times\\
    \times\int_{\cC_0}\prod_{i=1}^{N+2}\frac{\d z_i}{2\pi \i z_i^N}\;\frac{\prod_{1\leq i<j\leq N+2}\left(z_i-z_j\right)^2}{\prod_{i=1}^{N+2}\left(\bar{z}_1-z_i\right)^2 \left(\bar{z}_2-z_i\right)^2}\exp\left(-\frac{N}{t}\sum_{i=1}^{N+2}V\left(z_i\right)\right).
\end{multline}
Writing $U=\text{diag}\left(z_i\right)_{i=1,\dots,N+2}$, and noting that
\begin{align}
    \prod_{i=1}^{N+2}\left(\bar{z}_1-z_i\right)^2&=\det U^2\det\left(1-\bar{z}_1U^{-1}\right)^2,\\
    \prod_{i=1}^{N+2}\left(\bar{z}_2-z_i\right)^2&=\bar{z}_2^{2(N+2)}\det\left(1-\bar{z}_2^{-1}U\right)^2,
\end{align}
we find
\begin{multline}
    Z^{(1,1)}(N,t)=-\frac{(-1)^{(N+2)(N+1)/2}}{(N+2)!}\text{PV}\int_{\bar{\cC}_1^\star}\frac{\bar{z}_1^N\d\bar{z}_1}{2\pi \i}\int_{\bar{\cC}_2^\star}\frac{\bar{z}_2^N\d\bar{z}_2}{2\pi \i}\left(\bar{z}_1-\bar{z}_2\right)^2\exp\left(\frac{N}{t}\sum_{j=1}^2V\left(\bar{z}_j\right)\right)\times\\
    \times\int_{\cC_0}\prod_{i=1}^{N+2}\frac{\d z_i}{2\pi \i z_i^{N+2}}\;\frac{\prod_{1\leq i<j\leq N+2}\left(z_i-z_j\right)^2}{\det\left(1-\bar{z}_1U^{-1}\right)^2\cdot\bar{z}_2^{2(N+2)}\det\left(1-\bar{z}_2^{-1}U\right)^2}\exp\left(-\frac{N}{t}\sum_{i=1}^{N+2}V\left(z_i\right)\right).
\end{multline}
More compactly, we have
\begin{multline}
    Z^{(1,1)}(N,t)=-\text{PV}\int_{\bar{\cC}_1^\star}\frac{\d\bar{z}_1}{2\pi \i}\int_{\bar{\cC}_2^\star}\frac{\d\bar{z}_2}{2\pi \i}\;\bar{z}_1^N\bar{z}_2^{-N-4}\left(\bar{z}_1-\bar{z}_2\right)^2\exp\left(\frac{N}{t}\sum_{j=1}^2V\left(\bar{z}_j\right)\right)\times\\
    \times Z^{(0)}\left(N+2,t+2t/N\right)\ev{\frac{1}{\det\left(1-\bar{z}_1U^{-1}\right)^2\det\left(1-\bar{z}_2^{-1}U\right)^2}}_{(N+2,t+2t/N)}.
\end{multline}
We can also express the inverse determinants as exponentials of traces, and use the relations derived in appendix \ref{app} to obtain:
\begin{align}
    \ev{\frac{1}{\det\left(1-\bar{z}_1U^{-1}\right)^2\det\left(1-\bar{z}_2^{-1}U\right)^2}}&=\ev{\exp\left[-2\Tr\log\left(1-\bar{z}_1U^{-1}\right)-2\Tr\log\left(1-\bar{z}_2^{-1}U\right)\right]}\nonumber\\
    & \hspace{-2.2in} \approx\exp\Big[-2\ev{\Tr\log\left(1-\bar{z}_1U^{-1}\right)+\Tr\log\left(1-\bar{z}_2^{-1}U\right)}
    +4\ev{\Tr\log\left(1-\bar{z}_1U^{-1}\right)\Tr\log\left(1-\bar{z}_2^{-1}U\right)}_c\Big]\nonumber\\
    & \hspace{-2.2in} =\exp\left[\frac{N}{t}\left(\bar{z}_1+\frac{1}{\bar{z}_2}\right)-4\log\left(1-\frac{\bar{z}_1}{\bar{z}_2}\right)\right],
\end{align}
where, in the second line, we used the identity 
$\ev{e^{X}} = \exp \left[ \ev{X} + \frac{1}{2} \ev{X X}_c + \ldots \right]$.
The previous equation, along with the expressions for the effective potential inside and outside the unit circle, which we reproduce here for convenience,
\begin{align}
    \veff^{+}\left(\bar{z}_2\right) &= - \frac{\bar{z}_2}{2} + \frac{1}{2\bar{z}_2} - t \log (-\bar{z}_2) ,  \\
    \veff^{-}\left(\bar{z}_1\right) &= \frac{\bar{z}_1}{2} - \frac{1}{2\bar{z}_1} + t \log (-\bar{z}_1) ,
\end{align}
and the fact that $Z^{(0)}(N+2,t+2t/N)=Z^{(0)}(N,t)$ allow us to write:\footnote{Note that the regularization prescription no longer plays a role since the contours for $\bar{z}_1$ and $\bar{z}_2$ do not intersect.}
\begin{equation}
    \frac{Z^{(1,1)}(N,t)}{Z^{(0)}(N,t)}=-\int_{\bar{\cC}_1^\star}\frac{\d\bar{z}_1}{2\pi \i}\int_{\bar{\cC}_2^\star}\frac{\d\bar{z}_2}{2\pi \i}\;\frac{1}{\left(\bar{z}_1-\bar{z}_2\right)^2}e^{\frac{N}{t}\left[\veff^-\left(\bar{z}_1\right)+\veff^+\left(\bar{z}_2\right)\right]}.
\end{equation}
This expression is the analog of equation (\ref{Z11_Veff}) which was obtained in the description where the ghost instanton contributions are obtained by tunneling regular eigenvalues to the unphysical sheet. 
We can evaluate the integrals using the saddle-point approximation around $\left(\bar{z}_1,\bar{z}_2\right)=\left(z_1^\star,z_2^\star\right)$. Note that since both $\partial_z^2\veff^-\left(z_1^\star\right)$ and $\partial_z^2\veff^+\left(z_2^\star\right)$ are positive, the contours for $\delta \bar{z}_i=\bar{z}_i-z_i^\star$ will be along the imaginary axis, so each integral will contribute an extra factor of $\i$. 
Recalling that
\begin{equation}
    \frac{N}{t}\veff^-\left(z_1^\star\right)=\frac{N}{t}\veff^+\left(z_2^\star\right)=-NS_\text{strong}(t),
\end{equation}
we obtain
\begin{align}
    \frac{Z^{(1,1)}(N,t)}{Z^{(0)}(N,t)}&\approx-\frac{t}{2\pi N}\frac{1}{\left\vert\partial^2\veff^-\left(z_1^\star\right)\right\vert^{1/2}
    \left\vert\partial^2\veff^+\left(z_2^\star\right)\right\vert^{1/2}}\frac{1}{\left(z_1^\star-z_2^\star\right)^2}e^{-2NS_\text{strong}(t)}\\
    &=-\frac{t}{8\pi N\left(t^2-1\right)^{3/2}}e^{-2NS_{\text{strong}}(t)}.
\end{align}
This agrees with the result obtained in section \ref{sec:eiginst} and provides a derivation of the leading nonperturbative contribution to the GWW integral in the strong-coupling phase realized via anti-eigenvalue tunneling on the physical sheet.
\section{Unitary matrix integrals with general single-trace potentials}\label{appC}
The goal of this appendix is to extend the calculation of the leading (ghost) instanton contributions in the ungapped (strong-coupling) phase to the case of a matrix integral with a general single-trace potential:
\begin{equation}
    Z\left(N,t_k^{\pm}\right):=\int\frac{\d U}{\vol U(N)}\exp\left(\sum_{k=1}^K\left(\frac{t_k^+}{k}\Tr\left(U^k\right)+\frac{t_k^-}{k}\Tr\left(U^{-k}\right)\right)\right). \label{generalGWW}
\end{equation}
Here, the integral is over $N\times N$ unitary matrices $U$, the parameters $t_k^\pm$ are complex conjugates of one another, $t_k^+=\left(t_k^-\right)^\dagger$, and $K$ is an arbitrary positive integer. To guarantee that the general integral indeed admits an ungapped phase, where the distribution of eigenvalues in the large-$N$ limit covers the entire unit circle, we will impose the sufficient condition (see later discussion)
\begin{equation}
    \frac{2}{N}\sum_{k=1}^K\vert t_k^+\vert<1\, .\label{general-cond}
\end{equation}
The GWW integral is a special case where $K=1$ and
\begin{equation}
    t_1^+=t_1^-=\frac{N}{2t}\, .
\end{equation}

\subsection{Preliminaries}
Before discussing instanton corrections to the large-$N$ limit of this matrix integral, we will first note some useful results similar to those listed in appendix \ref{apperturbative}. Using the theorem of Diaconis and Shahshahani, equation (\ref{diacshah}), which we recall here for convenience,
\begin{align}
    \int \frac{\d U}{\vol U(N)} \, \prod_{j=1}^k 
    (\Tr U^j)^{a_j} (\Tr U^{-j})^{b_j} &= \prod_{j=1}^k j^{a_j} (a_j)!\, \delta_{a_j, b_j} 
    \quad\quad \text{if } N \geq \sum_{j=1}^k j \, a_j \, .
\end{align}
we note that the large-$N$ limit of the matrix integral (\ref{generalGWW}) evaluates to
\begin{align}
    Z_\infty\left(t_k^\pm\right)&=\int\frac{\d U}{\vol U(N)}\prod_{k=1}^K\left[\sum_{m=0}^\infty\sum_{n=0}^\infty\frac{1}{m!\, n!}\left(\frac{t_k^+}{k}\right)^m\left(\frac{t_k^-}{k}\right)^n\left(\Tr U^k\right)^m\left(\Tr U^{-k}\right)^n\right]\nonumber\\
    &=\prod_{k=1}^K\left[\sum_{m=0}^\infty\frac{1}{m!}\left(\frac{t_k^+t_k^-}{k}\right)^m\right]\nonumber\\
    &=\prod_{k=1}^K\exp\left(\frac{t_k^+t_k^-}{k}\right). \label{large-N-general}
\end{align}

The correlator we will need for the instanton computation is the same as the one in equation (\ref{eq:ev_exp_trlog_1}) which we recall here for convenience:
\begin{multline}
\label{eq:ev_exp_trlog_1-general}
    \ev{ e^{\Tr \log \big[ (1 - z_1^{-1} U )^2 (1 - z_2 U^{-1})^2 \big] } }
    \approx \exp \Big[ 2 \ev{\Tr \log (1 - z_1^{-1} U )} + 2 \ev{\Tr \log (1 - z_2 U^{-1} )} \\
    + 2 \ev{\Tr \log (1 - z_1^{-1} U ) \Tr \log (1 - z_1^{-1} U )}_c + 2 \ev{\Tr \log (1 - z_2 U^{-1} ) \Tr \log (1 - z_2 U^{-1} )}_c \\
    + 4 \ev{\Tr \log (1 - z_1^{-1} U ) \Tr \log (1 - z_2 U^{-1} )}_c\Big] \, .
\end{multline}
Similar to equation (\ref{singletrace-uj}), the single-trace large-$N$ correlator evaluates to
\begin{equation}
    \ev{\Tr U^j} = 
    \begin{cases}
        N \, , & j = 0 \, , \\
        t_j^- \, , & 1\leq j \leq K \, , \\
        t_{\vert j\vert}^+ \, , & -K\leq j \leq -1,\\
        0\, , & \text{otherwise.}
    \end{cases}
    \label{singletrace-uj-general}
\end{equation}
This correlator can be obtained immediately by taking derivatives of equations (\ref{generalGWW}) and (\ref{large-N-general}) with respect to the corresponding parameters $t_j^\pm$. Similarly, one can obtain the double-trace connected correlator
\begin{equation}
    \ev{\Tr U^j \Tr U^k}_c = \vert j\vert \, \delta_{j,-k} \, ,
\end{equation}
by noting that for $j,k\geq1$,
\begin{align}
    \ev{\Tr U^j \Tr U^k}_c&=\frac{jk}{Z_\infty}\partial_{t_j^+}\partial_{t_k^+}Z_\infty-\left(\frac{j}{Z_\infty}\partial_{t_j^+}Z_\infty\right)\left(\frac{k}{Z_\infty}\partial_{t_k^+}Z_\infty\right) = 0 \, ,\\
    \ev{\Tr U^{-j} \Tr U^{-k}}_c&=\frac{jk}{Z_\infty}\partial_{t_j^-}\partial_{t_k^-}Z_\infty-\left(\frac{j}{Z_\infty}\partial_{t_j^-}Z_\infty\right)\left(\frac{k}{Z_\infty}\partial_{t_k^-}Z_\infty\right) = 0 \, , \\
    \ev{\Tr U^j \Tr U^{-k}}_c&=\frac{jk}{Z_\infty}\partial_{t_j^+}\partial_{t_k^-}Z_\infty-\left(\frac{j}{Z_\infty}\partial_{t_j^+}Z_\infty\right)\left(\frac{k}{Z_\infty}\partial_{t_k^-}Z_\infty\right) = j\,\delta_{j,k} \, .
\end{align}

The single-trace correlator expression allows us to compute the single-log correlators
\begin{align}
    \ev{\Tr \log (1 - z U^{-1}) }
    &= - \sum_{j=1}^{\infty} \frac{1}{j} z^j \ev{\Tr U^{-j}} = -\sum_{j=1}^K\frac{1}{j}z^jt_j^+\, , 
    & |z| < 1 \, , \label{trlog1-general} \\
    \ev{\Tr \log (1 - z^{-1} U) }
    &=  - \sum_{j=1}^{\infty} \frac{1}{j} z^{-j} \ev{\Tr U^{j}} = -\sum_{j=1}^K\frac{1}{j}z^{-j}t_j^-\, , 
    & |z| > 1 \, . \label{trlog2-general}
\end{align}
The double-trace correlator allows us to compute the double-log connected correlators,
\begin{align}
    \ev{\Tr \log (1 - z_1 U^{-1}) \Tr \log (1 - z_2 U^{-1}) }_c
    &= \sum_{j,k=1}^{\infty} \frac{1}{jk} z_1^j z_2^k \ev{\Tr U^{-j} \Tr U^{-k} }_c =  0 \, ,
    & |z_1|,|z_2|<1 \, ,  \label{trlog3-general}\\
    \ev{\Tr \log (1 - z_1^{-1} U) \Tr \log (1 - z_2^{-1} U) }_c
    &= \sum_{j,k=1}^{\infty} \frac{1}{jk} z_1^{-j} z_2^{-k} \ev{\Tr U^{j} \Tr U^{k} }_c =  0 \, ,
    & |z_1|,|z_2| > 1 \, , \label{trlog4-general}\\
    \ev{\Tr \log (1 - z_1^{-1} U) \Tr \log (1 - z_2 U^{-1}) }_c
    &= \sum_{j,k=1}^{\infty} \frac{1}{jk} z_2^{j} z_1^{-k} \ev{\Tr U^{-j} \Tr U^{k} }_c 
    \nonumber \\
    &= - \log ( 1 - z_2/z_1) \, , 
    \quad\quad\quad
    \vert z_1\vert>1 \, ,\;\vert z_2\vert < 1 \, . \label{trlog5-general}
\end{align}
Substituting the results (\ref{trlog1-general})-(\ref{trlog5-general}) into (\ref{eq:ev_exp_trlog_1-general}), we get the following final expression for the correlator we will need:
\begin{equation}
    \label{eq:ev_exp_trlog_2-general}
    \ev{ e^{\Tr \log \big[ (1 - z_1^{-1} U )^2 (1 - z_2 U^{-1})^2 \big] } } \approx\exp\left[-2\sum_{j=1}^K\frac{1}{j}z_1^{-j}t_j^--2\sum_{j=1}^K\frac{1}{j}z_2^jt_j^+-4\log\left(1-z_2/z_1\right)\right] \, .
\end{equation}

\subsection{Resolvents, effective potentials, and the eigenvalue distribution}

We can now evaluate the resolvents, the effective potentials inside and outside the unit circle, as well as the eigenvalue distribution in the ungapped phase of the general matrix integral (\ref{generalGWW}). As in the case of the GWW integral, there will be two resolvents corresponding to the outside and inside of the unit circle, respectively,
\begin{align}
    R^+(z) &:= 
    \frac{1}{N} \left\langle \Tr \frac{1}{z-U} \right\rangle \, ,
    \quad \quad\text{for } \vert z \vert > 1 \label{rplusdef-general} \, , \quad \text{and} \\
    R^-(z) &:= \frac{1}{N} \left\langle \Tr \frac{1}{z-U} \right\rangle \, ,
    \quad \quad \text{for } \vert z \vert < 1 \, . \label{rminusdef-general}
\end{align}
We can evaluate these just as before, this time using (\ref{singletrace-uj-general}),
\begin{align}
    R^+(z) &=\frac{1}{N}\sum_{k=1}^\infty z^{-k}\ev{\Tr U^{k-1}}=\frac{1}{z}+\frac{1}{N}\sum_{k=1}^K t_k^- z^{-k-1}\, ,\\
    R^-(z) &=-\frac{1}{N}\sum_{k=1}^\infty z^{k-1}\ev{\Tr U^{-k}}=-\frac{1}{N}\sum_{k=1}^K t_k^+ z^{k-1}\, .
\end{align}
The density of states in the general matrix integral (\ref{generalGWW}) will be given by
\begin{equation}
    \rho(\theta)=\frac{R^+(z)-R^-(z)}{2\pi \i}\frac{\d z}{\d\theta}=\frac{1}{2\pi}\left(1+\frac{1}{N}\sum_{k=1}^K\left(t_k^+e^{ik\theta}+t_k^-e^{-ik\theta}\right)\right)\, \label{density-general} .
\end{equation}
We can check that the condition (\ref{general-cond}) ensures the eigenvalue density is non-negative. Writing
\begin{equation}
    t_k^\pm=\frac{N}{2q_k}e^{\pm i\varphi_k}\, ,
\end{equation}
where $q_k>0$ and $\varphi_k\in[0,2\pi)$, the condition (\ref{general-cond}) takes the form
\begin{equation}
    \sum_{k=1}^K\frac{1}{q_k}<1\; ,
\end{equation}
and the eigenvalue density takes the form
\begin{equation}
    \rho(\theta)=\frac{1}{2\pi}\left(1+\sum_{k=1}^K\frac{1}{q_k}\cos\left(k\theta+\varphi_k\right)\right)\, .
\end{equation}
Thus, $\rho(\theta)>0$ for all $\theta\in[0,2\pi)$.

We can now evaluate the derivative of the effective potential inside and outside the unit circle. For convenience, we will modify the conventions used in section \ref{sec:eiginst} slightly. We adapt the holomorphic form of the matrix integral, equations (\ref{zntholomorphic1}) and (\ref{zntholomorphic}), to the present case:
\begin{align}
    Z\left(N,t_k^\pm\right) &= (-1)^{\frac{1}{2}N(N-1)}\frac{1}{N!} \int 
    \prod_{i=1}^N \frac{\d z_i}{2\pi \i z_i^N} \prod_{j<k} (z_j - z_k)^2
    \exp \left( -\sum_{i=1}^N V(z_i) \right) \label{zntholomorphic1-general} \\
    &=(-1)^{\frac{1}{2}N(N-1)} \, \i^{-N}\frac{1}{N!} \int 
    \prod_{i=1}^N \frac{\d z_i}{2\pi} \prod_{j<k} (z_j - z_k)^2
    \exp \left( -N\sum_{i=1}^N \left(\frac{1}{N}V\left(z_i\right) + \log z_i\right) \right) \, ,
    \label{zntholomorphic-general}
\end{align}
where
\begin{equation}
    V(z)=-\sum_{k=1}^K\left(\frac{t_k^+}{k}z^k+\frac{t_k^-}{k}z^{-k}\right)
\end{equation}
The derivative of the effective potential is then obtained as
\begin{align}
    \frac{\d}{\d z}\veff^+(z) &= \frac{1}{N}V'\left(z\right)+\frac{1}{z}-2R^+(z) \nonumber\\
    &= -\frac{1}{N}\sum_{k=1}^K\left(t_k^+z^{k-1}-t_k^-z^{-k-1}\right)+\frac{1}{z}-2\left(\frac{1}{z}+\frac{1}{N}\sum_{k=1}^K t_{k}^- z^{-k-1}\right) \nonumber\\
    &=-\frac{1}{z}-\frac{1}{N}\sum_{k=1}^K\left(t_k^+z^{k-1}+t_k^-z^{-k-1}\right)\, \label{eqn:dveff+-general} ,\\
    \frac{\d}{\d z}\veff^-(z) &=\frac{1}{N}V'(z)+\frac{1}{z}-2R^-(z)\nonumber\\
    &=-\frac{1}{N}\sum_{k=1}^K\left(t_k^+z^{k-1}-t_k^-z^{-k-1}\right)+\frac{1}{z}+\frac{2}{N}\sum_{k=1}^Kt_k^+z^{k-1}\nonumber\\
    &=\frac{1}{z}+\frac{1}{N}\sum_{k=1}^K\left(t_k^+z^{k-1}+t_k^-z^{-k-1}\right)\, \label{eqn:dveff--general}.
\end{align}
As in the GWW case, we see that the analytic continuations of the derivatives of the effective potential are additive inverses of each other. Furthermore, upon integrating,
\begin{align}
    \veff^+(z) &= -\log z-\frac{1}{N}\sum_{k=1}^K\left(\frac{t_k^+}{k}z^k-\frac{t_k^-}{k}z^{-k}\right)\, \label{veff+gen},\\
    \veff^-(z) &=\log z+\frac{1}{N}\sum_{k=1}^K\left(\frac{t_k^+}{k}z^k-\frac{t_k^-}{k}z^{-k}\right)\, \label{veff-gen},
\end{align}
we note that
\begin{equation}
    \veff^-\left(1/z^\dagger\right)=\left(\veff^+(z)\right)^\dagger\, ,\label{relationVeff}
\end{equation}
where the dagger symbol denotes complex conjugation.
The eigenvalue instanton locations are given by the common zeros of (\ref{eqn:dveff+-general}) and (\ref{eqn:dveff--general}),
\begin{equation}
    N + \sum_{k=1}^K \left(t_k^+\left(z^\star\right)^k + t_k^-\left(z^\star\right)^{-k}\right) = 0 \, . 
    \label{polynomial-roots}
\end{equation}
This equation can be written as a polynomial equation of degree $2K$ and therefore has $2K$ complex solutions. 
However, an important observation is that if a complex number $z^\star$ is a zero of (\ref{polynomial-roots}) then $1/\left(z^\star\right)^\dagger$ is also a zero of it.
This can be seen immediately by taking the complex conjugate of (\ref{polynomial-roots}). 
Thus, equation (\ref{polynomial-roots}) has $K$ pairs of zeroes, $z^\star$ and $1/\left(z^\star\right)^\dagger$, where one element of each pair is inside the unit circle and the other is outside. Note that there are no solutions on the unit circle itself since the left side of equation (\ref{polynomial-roots}) is proportional to the eigenvalue density (\ref{density-general}), which is nowhere vanishing due to condition (\ref{general-cond}).

Thus, like in the GWW case, the eigenvalue instantons come in pairs, whose leading contribution we evaluate in the next subsection.
\subsection{Leading-order two-instanton contributions}
We are interested in obtaining the analog of (\ref{eqn:11contribution}) associated to the eigenvalue instanton locations $z^\star$ and $1/\left(z^\star\right)^\dagger$ for the matrix integral with a general single-trace potential, (\ref{generalGWW}). Our starting point is the analog of equation (\ref{eqn:Z11start}),
\begin{multline}
    Z^{(1,1)}\left(N,t_k^\pm\right) = \frac{1}{(N-2)!} (-1)^{N(N-1)/2} \int_{\cC_1} \frac{\d z_1}{2\pi\i z_1^N} e^{-V(z_1)} \int_{\cC_2} \frac{\d z_2}{2\pi\i z_2^N} e^{-V(z_2)} (z_1 - z_2)^2 \\
    \times \int_{\cC_0} \prod_{i=3}^{N} \frac{\d z_i}{2\pi\i z_i^N} 
    e^{-\sum_{i=3}^{N} V(z_i)} 
    \times \prod_{j=3}^{N} \big[ (z_1 - z_j)^2 (z_2 - z_j)^2 \big] \times \prod_{3\leq k < l \leq N} (z_k - z_l)^2 \, ,
\end{multline}
where we once again start by considering $\vert z_1 \vert > 1$ and $ \vert z_2 \vert < 1$. We now follow the same steps laid out in equations (\ref{eqn:222}) - (\ref{eqn:11contribution}). First rewrite the previous equation as:
\begin{multline}
    Z^{(1,1)}\left(N,t_k^\pm\right) = \frac{1}{(N-2)!} (-1)^{N(N-1)/2} \int_{\cC_1} \frac{\d z_1}{2\pi\i z_1^N} e^{-V(z_1)} \int_{\cC_2} \frac{\d z_2}{2\pi\i z_2^N} e^{-V(z_2)} (z_1 - z_2)^2 \\
    \times z_1^{2(N-2)} \int_{\cC_0} \prod_{i=3}^{N} \frac{\d z_i}{2\pi\i z_i^{N-2}} e^{-\sum_{i=3}^{N}  V(z_i)} \times \prod_{j=3}^{N} \big[ (1 - z_j/z_1)^2 (1 - z_2/z_j)^2 \big] \times \prod_{3\leq k < l \leq N} (z_k - z_l)^2 \, .
\end{multline}
This allows us to perform the integrals in $z_3,\dots,z_N$ and express the result as a correlator in the analog of (\ref{generalGWW}):
\begin{equation}
\begin{split}
    Z^{(1,1)}\left(N,t_k^\pm\right) 
    &= - \int_{\cC_1} \frac{\d z_1}{2\pi\i} \int_{\cC_2} \frac{\d z_2}{2\pi\i} e^{-V(z_1) - V(z_2)} \times \frac{z_1^{N-4} (z_1 - z_2)^2 }{z_2^N} \\
    & \hspace{0.75in} \times Z^{(0)}\left(N-2,t_k^\pm\right) \ev{ e^{\Tr \log \big[ (1 - z_1^{-1} U )^2 (1 - z_2 U^{-1})^2 \big] } }_{(N-2,t_k^\pm)} \\
    &= - \int_{\cC_1} \frac{\d z_1}{2\pi\i} \int_{\cC_2} \frac{\d z_2}{2\pi\i} e^{-V(z_1) - V(z_2)} \times \frac{z_1^{N-4} (z_1 - z_2)^2 }{z_2^N} \\
    & \hspace{0.75in} \times Z^{(0)}\left(N,t_k^\pm\right) \exp \left[-2\sum_{k=1}^K\frac{1}{k}z_1^{-k}t_k^--2\sum_{k=1}^K\frac{1}{k}z_2^kt_k^+-4\log\left(1-z_2/z_1\right)\right] \, ,
\end{split}
\end{equation}
which follows since in the large-$N$ limit, $Z^{(0)}\left(N-2,t_k^\pm\right)=Z^{(0)}\left(N,t_k^\pm\right)$, and using the expression for the correlator (\ref{eq:ev_exp_trlog_2-general}). The previous relation has a simple form in terms of the effective potentials, which is analogous to (\ref{Z11_Veff}):
\begin{equation}
\label{Z11_Veff-gen}
    \frac{Z^{(1,1)}\left(N,t_k^\pm\right)}{Z^{(0)}\left(N,t_k^\pm\right)} = - \int_{\cC_1} \frac{\d z_1}{2\pi\i} \int_{\cC_2} \frac{\d z_2}{2\pi\i} e^{-N \big[ \veff^{+}(z_1) + \veff^{-}(z_2) \big]} \frac{1}{(z_1 - z_2)^2} \, .
\end{equation}
There are now two possibilities depending on whether we expect a regular eigenvalue instanton or a ghost instanton contribution. These depend on whether we choose $\cC_1$ to pass through $z^\star$ and $\cC_2$ to pass through $1/\left(z^\star\right)^\dagger$ or vice versa. 
For concreteness, let us assume without loss of generality that $z^\star$ is the eigenvalue instanton location outside the unit circle and $1/\left(z^\star\right)^\dagger$ is inside the unit circle.

Taking $\cC_1$ to pass through $z^\star$ and $\cC_2$ to pass through $1/\left(z^\star\right)^\dagger$, we obtain the regular eigenvalue instanton contribution:
\begin{equation}
    \frac{Z^{(1,1)}\left(N,t_k^\pm\right)}{Z^{(0)}\left(N,t_k^\pm\right)} = \frac{1}{2\pi N} \frac{1}{\sqrt{\partial^2 \veff^{+}(z^\star)}\sqrt{\partial^2 \veff^{-}\left(1/\left(z^\star\right)^\dagger\right)}} \frac{1}{\left(z^\star - 1/\left(z^\star\right)^\dagger\right)^2} e^{ - 2 NS_{\text{strong}}^{\text{regular}}}\, ,
\label{eqn:11contribution-gen-regular}
\end{equation}
where
\begin{equation}
    S_{\text{strong}}^{\text{regular}}:=\frac{1}{2}\left[\veff^+\left(z^\star\right)+\veff^-\left(1/\left(z^\star\right)^\dagger\right)\right]\, .\label{action-regular-gen}
\end{equation}
Note that $S_{\text{strong}}^{\text{regular}}$ is real due to the relation (\ref{relationVeff}) between the effective potentials. The regular eigenvalue instanton pair contribution is the one that determines the leading nonperturbative correction associated to the pair $\left(z^\star,\,1/\left(z^\star\right)^\dagger\right)$ to the general matrix integral (\ref{generalGWW}) when $S_{\text{strong}}^{\text{regular}}>0$.

Taking $\cC_1$ to pass through $1/\left(z^\star\right)^\dagger$ and $\cC_2$ to pass through $z^\star$, we obtain the ghost instanton contribution:
\begin{equation}
    \frac{Z^{(1,1)}\left(N,t_k^\pm\right)}{Z^{(0)}\left(N,t_k^\pm\right)} = \frac{1}{2\pi N} \frac{1}{\sqrt{\partial^2 \veff^{+}\left(1/\left(z^\star\right)^\dagger\right)}\sqrt{\partial^2 \veff^{-}\left(z^\star\right)}} \frac{1}{\left(z^\star - 1/\left(z^\star\right)^\dagger\right)^2} e^{ - 2 NS_{\text{strong}}^{\text{ghost}}}\, ,
\label{eqn:11contribution-gen-ghost}
\end{equation}
where
\begin{equation}
    S_{\text{strong}}^{\text{ghost}}:=\frac{1}{2}\left[\veff^+\left(1/\left(z^\star\right)^\dagger\right)+\veff^-\left(z^\star\right)\right]\, .\label{action-ghost-gen}
\end{equation}
Similarly, $S_{\text{strong}}^{\text{ghost}}$ is real due to relation (\ref{relationVeff}) between the effective potentials. Note also that instanton action corresponding to the ghost instanton is the additive inverse of the instanton action corresponding to the regular instanton,
\begin{equation}
    S_{\text{strong}}^{\text{ghost}}=-S_{\text{strong}}^{\text{regular}}.
\end{equation}
The ghost instanton pair contribution is the one that determines the leading nonperturbative correction associated to the pair $\left(z^\star,\,1/\left(z^\star\right)^\dagger\right)$ to the general matrix integral (\ref{generalGWW}) when $S_{\text{strong}}^{\text{ghost}}>0$.

Note that the leading regular eigenvalue instanton contribution (\ref{eqn:11contribution-gen-regular}) differs from the leading ghost instanton contribution (\ref{eqn:11contribution-gen-ghost}) only in the sign of the action $S_{\text{strong}}$ and an overall sign due to the relations between the effective potentials.

This method also allows us to compute the contributions from other, more general, instanton configurations.
The defining contour of the matrix integral tells us how to weigh these various contributions.

\subsection{The special case of the GWW matrix integral}

Finally, we show how the leading-order ghost-instanton contribution (\ref{eqn:11contribution-gen-ghost}) in the matrix integral with a general single-trace potential reduces to its counterpart (\ref{eqn:11contribution}) in the GWW matrix integral. 
The GWW integral (\ref{defznt}) is a special case of the general matrix integral (\ref{generalGWW}) where $K=1$ and $t_1^+=t_1^-=\frac{N}{2t}$. The effective potentials (\ref{veff+gen}) and (\ref{veff-gen}) reduce to
\begin{align}
    \veff^+(z)=-\log z-\frac{1}{2t}\left(z-\frac{1}{z}\right)\, ,\\
    \veff^-(z)=\log z+\frac{1}{2t}\left(z-\frac{1}{z}\right)\, ,
\end{align}
and the eigenvalue instanton location condition (\ref{polynomial-roots}) reduces to
\begin{equation}
    1+\frac{1}{2t}\left(\left(z^\star\right)+\frac{1}{z^\star}\right)=0\, ,
\end{equation}
which has solutions
\begin{equation}
    z^\star=-t-\sqrt{t^2-1}\, ,\qquad\qquad 1/\left(z^\star\right)^\dagger=-t+\sqrt{t^2-1}\,.
\end{equation}
The regular instanton action (\ref{action-regular-gen}) reduces to:
\begin{equation}
    S_{\text{strong}}^{\text{regular}}=-\left(\log\left(t+\sqrt{t^2-1}\right)-\sqrt{1-t^{-2}}\right)\, ,
\end{equation}
while the ghost instanton action (\ref{action-ghost-gen}) reduces to:
\begin{equation}
    S_{\text{strong}}^{\text{ghost}}=+\left(\log\left(t+\sqrt{t^2-1}\right)-\sqrt{1-t^{-2}}\right)\, .
\end{equation}
Thus, the GWW matrix integral will have a leading ghost-instanton contribution in the strong-coupling phase. Additionally,
\begin{align}
    \partial^2\veff^+\left(1/\left(z^\star\right)^\dagger\right) &= -\left(z^\star\right)^2\sqrt{1-t^{-2}}\, ,\\
    \partial^2\veff^-\left(z^\star\right) &= -\frac{1}{\left(z^\star\right)^2}\sqrt{1-t^{-2}}\, ,\\
    \left(z^\star-1/\left(z^\star\right)^\dagger\right)^2 &= 4\left(t^2-1\right)
\end{align}
Therefore, the leading nonperturbative contribution to the large-$N$ limit of the GWW matrix integral in the strong-coupling phase will be a special case of (\ref{eqn:11contribution-gen-ghost}),
\begin{equation}
    \frac{Z^{\left(1,1\right)}\left(N,t_k^\pm\right)}{Z^{(0)}\left(N,t_k^\pm\right)}=-\frac{t}{8\pi N\left(t^2-1\right)^{3/2}}e^{-2NS_{\text{strong}}^{\text{ghost}}}\,.
\end{equation}
This agrees with (\ref{eqn:11contribution}).
\bibliographystyle{apsrev4-1long}
\bibliography{main}
\end{document}